\documentclass[12pt]{iopartfn}

\usepackage{iopams}  
\usepackage{color}
\usepackage[dvipsnames]{xcolor} 
\usepackage{tikz}
\usepackage{psfrag}
\usepackage{graphicx}
\usepackage{stackrel}
\usepackage{amsfonts}
\usepackage{bm}          
\usepackage{amssymb}

\newcommand{\be}{\begin{equation}}
\newcommand{\ee}{\end{equation}}
\newcommand{\ba}{\begin{eqnarray}}
\newcommand{\ea}{\end{eqnarray}}

\def\N{{\mathbb N}}

\begin{document}

\title[Critical polynomials]{Transfer matrix computation of critical polynomials
for two-dimensional Potts models}

\author{Jesper Lykke Jacobsen$^{1,2}$ and Christian R.\ Scullard$^{3}$}
\address{${}^1$LPTENS, \'Ecole Normale Sup\'erieure, 24 rue Lhomond, 75231
Paris, France}
\address{${}^2$Universit\'e Pierre et Marie Curie, 4 place Jussieu, 75252 Paris,
France}
\address{${}^3$Livermore, California, USA}

\eads{\mailto{jesper.jacobsen@ens.fr}, \mailto{scullard1@llnl.gov}}

\begin{abstract}

  In our previous work \cite{Jacobsen12} we have shown that critical manifolds
of the $q$-state Potts model
  can be studied by means of a graph polynomial $P_B(q,v)$, henceforth referred
to as the critical
  polynomial. This polynomial may be defined on any periodic two-dimensional
lattice.
  It depends on a finite subgraph $B$, called the basis, and
  the manner in which $B$ is tiled to construct the lattice. The real
  roots $v = {\rm e}^K - 1$ of $P_B(q,v)$ either give the exact critical points
  for the lattice, or provide approximations that, in principle, can be made
arbitrarily accurate by increasing the size of $B$ in an appropriate way. In
earlier work, $P_B(q,v)$ was defined by a contraction-deletion identity, similar
to
  that satisfied by the Tutte polynomial. Here, we give a probabilistic
definition of $P_B(q,v)$, which facilitates its computation, using the transfer
matrix, on much larger $B$ than was previously possible.

  We present results for the critical polynomial on the $(4,8^2)$, kagome, 
  and $(3,12^2)$ lattices for bases of up to respectively 96, 162, and
  243 edges, compared to the limit of 36 edges with
  contraction-deletion. We discuss in detail the role of the
  symmetries and the embedding of $B$. The critical temperatures $v_c$ obtained
  for ferromagnetic ($v>0$) Potts models are at least as precise as the best
available results
  from Monte Carlo simulations or series expansions. For instance, with $q=3$ we
obtain
  $v_c(4,8^2) = 3.742\,489 \, (4)$, $v_c(\mathrm{kagome}) = 1.876\,459\,7
\,(2)$,
  and $v_c(3,12^2) = 5.033\,078\,49 \,(4)$, the precision being comparable or
superior to the best simulation results.
  More generally, we trace the critical manifolds in the real $(q,v)$ plane and
discuss
  the intricate structure of the phase diagram in the antiferromagnetic ($v<0$)
region.

\end{abstract}

\noindent

\section{Introduction}

The $q$-state Potts model \cite{Potts52} is one of
the most well-studied models of statistical physics \cite{Wu82,Baxter_book}.
Given a connected graph $G=(V,E)$ with vertex set $V$ and edge
set $E$, its partition function $Z$ is most conveniently expressed in the
Fortuin-Kasteleyn representation \cite{FK1972}
\be
 Z = \sum_{A \subseteq E} v^{|A|} q^{k(A)} \,,
 \label{FK_repr}
\ee
where $|A|$ denotes the number of edges in the subset $A$, and $k(A)$
is the number of connected components in the induced graph $G_A =
(V,A)$. The temperature parameter $v$ is related to the reduced interaction
energy $K$ between adjacent $q$-component spins through 
$v = {\rm e}^K - 1$. In the representation (\ref{FK_repr}), both $q$ and $v$
can formally be allowed to take arbitrary real values.

In two dimensions, the Potts model can in general only be exactly
solved along certain curves in the $(q,v)$ plane, and for a very few
regular lattices $G$. This includes the square \cite{Baxter73} and triangular
lattices \cite{Baxter78}, and its dual hexagonal lattice. The solution on the
triangular lattice can be extended by decoration \cite{Wu10,Scullard06,Ziff06}
and to the
closely related bowtie lattices \cite{Wierman84,Ding12,SJ12c}.
The critical manifold---which is the set of points in $(q,v)$ space at which the
model stands at a
phase transition---is obviously of special interest. Remarkably,
in the solvable cases \cite{Baxter73,Baxter78,Baxter82}, the loci of exact
solvability
coincide precisely with the critical manifold.  Moreover, the critical manifolds
are given by simple algebraic curves:
\ba
 (v^2-q)(v^2+4v+q) &=& 0 \,, \qquad \mbox{(square lattice)} 
 \label{sq_latt_cc} \\
 v^3 + 3v^2 - q    &=& 0 \,, \qquad \mbox{(triangular lattice)}
 \label{tri_latt_cc} \\
 v^3 - 3q v - q^2  &=& 0 \,. \qquad \, \mbox{(hexagonal lattice)}
 \label{hex_latt_cc}
\ea

The critical manifolds on other Archimedean lattices---such as the $(4,8^2)$,
kagome and $(3,12^2)$ lattices---are long-standing unsolved problems of
lattice statistics.
Recently, we introduced \cite{Jacobsen12} a graph polynomial
$P_B(q,v)$---henceforth
referred to as the critical polynomial---as a step towards solving such
problems.
This polynomial may be defined on any periodic two-dimensional lattice $G$.
It depends on a finite subgraph $B$, called the basis, and
the way in which the basis is tiled to form $G$. It turns out that in the
exactly solvable cases,
$P_B(q,v)$ factorises for any choice of $B$, shedding a small factor which is
precisely
given by Eqs.~(\ref{sq_latt_cc})--(\ref{hex_latt_cc}). In the unsolved cases,
$P_B(q,v)$ does not factorise, except for a few fortuitous choices of $B$, but
the real roots $v_c$ of $P_B(q,v)$ provide approximations to the critical
temperature
that become more accurate with appropriately increasing size of $B$.

The definition of $P_B(q,v)$ made in \cite{Jacobsen12} was through a
contraction-deletion identity, similar to
that satisfied by (\ref{FK_repr}), and enabled the practical computation of
$P_B(q,v)$ for bases of size
up to 36 edges (see also \cite{Scullard11-2} and \cite{Scullard12}). For the
kagome lattice it was found that the smallest possible 6-edge basis
reproduced a well-known, but now refuted \cite{ZiffSuding97}, conjecture by Wu \cite{Wu79}. From
comparisons with
high-precision numerical simulations, it was found that results for $v_c$ in the
ferromagnetic regime ($v>0$)
improved by two orders of magnitude when going from the 6-edge to the 36-edge
basis.

The purpose of this paper is threefold. First, we extend the field of
investigations to include
also the $(4,8^2)$ and $(3,12^2)$ lattices. Second, and more importantly, we
provide an
alternative probabilistic definition of $P_B(q,v)$, which allows for
much more efficient computations, by using the transfer matrix, than
was previously possible with contraction-deletion. The alternative definition
permits us to obtain
the critical polynomial on the $(4,8^2)$, kagome, 
and $(3,12^2)$ lattices for bases of up to respectively 96, 162, and 243 edges.
The improvement over \cite{Jacobsen12} is such that the precision on $v_c$ in
the ferromagnetic
regime is comparable or superior to that of the best available results using
alternative methods,
be it exact transfer matrix diagonalisations, Monte Carlo simulations or series
expansions.
For instance, with $q=3$ we obtain
\ba
  v_c(4,8^2) &=& 3.742\,489 \, (4) \,, \\
  v_c(\mathrm{kagome}) &=& 1.876\,459\,7 \,(2) \,, \\
  v_c(3,12^2) &=& 5.033\,078\,49 \,(4) \,,
\ea
where the number in parentheses indicates the error bar on the last quoted
digit.
The third purpose is to use the critical polynomials to trace the (very accurate
approximations to the)
critical manifolds in the real $(q,v)$ plane. This in particular reveals a very
intricate structure of the phase diagrams in the antiferromagnetic ($v<0$)
region.
 
The layout of the paper is as follows.
In section~\ref{sec:crit_pol}, after recalling the contraction-deletion
definition \cite{Jacobsen12}
of the critical polynomial $P_B(q,v)$, we present the alternative probabilistic
definition and give
some details on the bases $B$ to be considered. The latter definition opens the
possibility of
computing $P_B(q,v)$ from a transfer matrix construction, which is described in
section~\ref{sec:tm}.
The results for the $(4,8^2)$, kagome and $(3,12^2)$ lattices are presented in
section~\ref{sec:results},
where we also provide numerical values of the critical points $v_c$ in the
ferromagnetic regime.
The phase diagrams in the real $(q,v)$ plane are discussed in
section~\ref{sec:phase_diag}.
Finally, section~\ref{sec:disc} is dedicated to a discussion and further
perspectives.

\section{The critical polynomial}
\label{sec:crit_pol}

We illustrate the contraction-deletion definition \cite{Jacobsen12} of
$P_B(q,v)$ by
means of a specific example. First recall that the partition function
(\ref{FK_repr}) with general edge-dependent weights $\{v\}$ satisfies the 
contraction-deletion identity \cite{Sokal05}
\be
 Z_G(q,\{v\}) = v_e Z_{G/e}(q,\{v\}) + Z_{G\setminus e}(q,\{v\}) \,,
 \label{cont_del}
\ee
where $e \in E$ is any edge in $G$. Here $G/e$ denotes the graph obtained from
$G$ by contracting $e$ to
a point and identifying the vertices at its end points (if they are
different), and $G \setminus e$ denotes the graph obtained from $G$
by deleting $e$. 

\begin{figure}
\begin{center}
\includegraphics{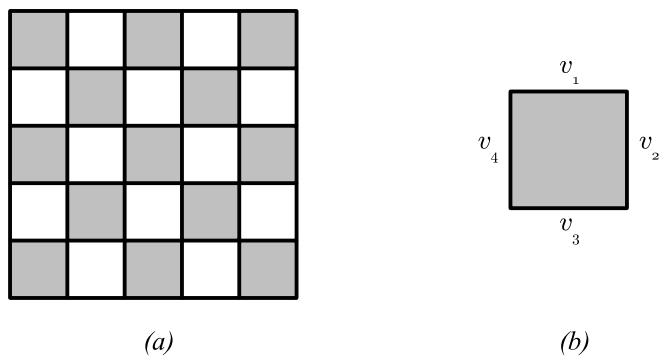}
\caption{a) The square lattice with the checkerboard couplings shown in b).}
\label{fig:checkerboard}
\end{center}
\end{figure}

Now let $G$ be the square lattice, and choose the 4-edge basis $B$
with couplings $\{v_1,v_2,v_3,v_4\}$ shown in Figure~\ref{fig:checkerboard}b. We
choose the checkerboard embedding of $B$ in $G$ shown in
Figure~\ref{fig:checkerboard}a. The contraction-deletion definition
\cite{Jacobsen12}
amounts to assuming that the critical polynomial $P_B(q,v)$ satisfies the same
identity as (\ref{cont_del}) for any edge $e \in B$.
Performing the deletion-contraction of the edge with weight $v_4$ we thus obtain
\be
 P_B(q,\{v_1,v_2,v_3,v_4\}) = v_4 P_{B^{\rm tri}}(q,\{v_1,v_2,v_3\})
  + P_{B^{\rm hex}}(q,\{v_1,v_2,v_3\}) \,.
 \label{PB_tri_hex}
\ee
In the first (contracted) term the embedded basis now spans the triangular
lattice,
so we can insert the known exact result
\be
 P_{B^{\rm tri}}(q,\{v_1,v_2,v_3\}) = v_1 v_2 v_3 + v_1 v_2 + v_2 v_3 + v_3 v_1
- q \,,
 \label{PB_tri}
\ee
that generalises (\ref{tri_latt_cc}). In the second (deleted) term we recover
the
hexagonal lattice, for which the exact result generalising (\ref{hex_latt_cc})
reads
\be
 P_{B^{\rm hex}}(q,\{v_1,v_2,v_3\}) = v_1 v_2 v_3 - q(v_1 + v_2 + v_3) - q^2 \,.
 \label{PB_hex}
\ee
Inserting (\ref{PB_tri})--(\ref{PB_hex}) into (\ref{PB_tri_hex}) we arrive at
the
desired critical polynomial
\ba
 P_B(q,\{v_1,v_2,v_3,v_4\}) &=& v_1 v_2 v_3 v_4 +
 v_2 v_3 v_4 + v_1 v_3 v_4 + v_1 v_2 v_4 + v_1 v_2 v_3 \nonumber \\
 & & -q (v_1 + v_2 + v_3 + v_4) - q^2 \,.
 \label{PB_check}
\ea
Note that the expression $P_B(q,\{v_1,v_2,v_3,v_4\}) = 0$ coincides with a
result
derived by Wu \cite{Wu79} using a different method (and a homogeneity
assumption).

Critical polynomials $P_B(q,v)$ defined in this way are unique, that is,
they are a property only of the basis $B$ and the way in which $B$ is
embedded in the infinite lattice $G$. In particular, $P_B(q,v)$ is
independent of the order in which edges are contracted-deleted
\cite{Scullard11-2}. 

In the particular case considered above, (\ref{PB_check}) actually provides
the exact critical manifold of the square lattice with checkerboard couplings \cite{SJ12c}.
However, as mentioned in the Introduction and discussed in details in
\cite{Jacobsen12},
in general we only recover an approximation to the critical manifold,
that converges to the true critical manifold upon letting the size of $B$ go to
infinity (at finite aspect ratio). How close one can get to $v_c$ is thus
limited by one's ability to
actually compute the polynomial on large $B$. In \cite{Jacobsen12}, a
computer program was used to perform the contraction-deletion
algorithm on various bases for the kagome lattice. However, this
algorithm is exponential in the number of edges in $B$, and the upper
limit of feasibility was $36$ edges.

\begin{figure}
\begin{center}
\includegraphics{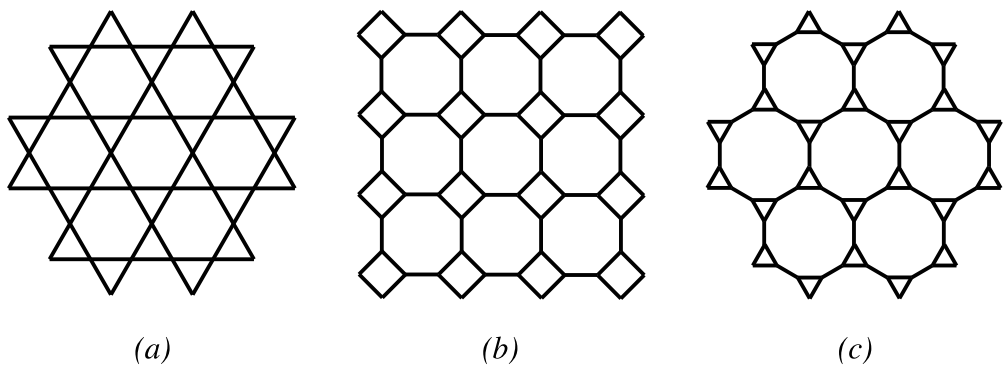}
\caption{a) the kagome lattice; b) the $(4,8^2)$ lattice; c) the $(3,12^2)$
lattice.}
\label{fig:bondlattices}
\end{center}
\end{figure}

Below, we present an alternative definition of $P_B(q,v)$ in terms of
probabilities of events on $B$. This permits use of a transfer matrix approach,
a much more efficient algorithm that is, roughly speaking, exponential only in
the number of vertices across a
horizontal cross-section of $B$.  By these means, we will be able to compute
critical polynomial on the $(4,8^2)$, kagome, 
and $(3,12^2)$ lattices for bases of up to respectively 96, 162, and 243 edges.
These three lattices are shown in Figure~\ref{fig:bondlattices}. We note that in parts of sections~\ref{sec:crit_pol}--\ref{sec:tm} we will present material that has been discussed for $q=1$ in our previous work on percolation \cite{SJ12}. Many of these concepts are identical or only slightly modified in the Potts generalization, but for the sake of completeness we explain these in full below, reproducing some passages verbatim from \cite{SJ12}.

\subsection{Alternative definition}
\label{sec:altdef}

According to (\ref{FK_repr}), the probability of any event on the finite graph
$B$ is proportional to a sum of terms of the type $\left( \prod_{i \in A} v_i
\right) q^{k(A)}$,
where $A$ are some subsets of the edges in $B$
describing which edges need to be present in order to realise the
event. We are here interested in the probabilistic, geometrical
interpretation of the critical polynomials $P_B(q,v)$ in terms of such events.
But to discuss
this, we will first need some definitions.

The infinite lattice $G$ is partitioned into identical subgraphs $B$,
and we assume that each is in the same edge-state. We are interested in the
global connectivity
properties of the system. If, given any two copies of the basis, $B_1$
and $B_2$, separated by an arbitrary distance, it is possible to
travel from $B_1$ to $B_2$ along an open path, then we say that there
is an infinite two-dimensional (2D) cluster in the system. We denote
the weight of this event $W(2D;B)$. On the other hand, if it is
not possible to connect any non-neighbouring $B_1$ and $B_2$, then
there are no infinite clusters in the system, a situation whose
weight we write as $W(0D;B)$. The third possibility is that {\it
  some} arbitrarily separated $B_1$ and $B_2$ are connected, but not
all, indicating the presence of infinite one-dimensional (1D) paths
(or filaments), and we denote the corresponding weight
$W(1D;B)$.

We have found that all the (inhomogeneous) critical polynomials
$P_B(q,\{v\})$ that we have computed using the contraction-deletion
definition%
\footnote{Examples include, but are not limited to, all the cases discussed in
\cite{Jacobsen12}.}
can be rewritten very simply as
\begin{equation}
 P_B(q,\{v\}) = W(2D;B) - q \, W(0D;B) \,.
 \label{eq:2D0D}
\end{equation}
Despite its apparent simplicity, eq.~(\ref{eq:2D0D}) is the main
result of this paper. 

To make completely clear the meaning of (\ref{eq:2D0D}) we need to discuss
two important points.

First, one may wish to think of the quantities $W(2D;B)$
and $W(0D;B)$ appearing on the left-hand side of (\ref{eq:2D0D}) as describing
the {\em probabilities} of the events defined above. However, since the critical
manifold
is found from $P_B(q,\{v\}) = 0$, normalisation issues are not important. It is
thus more convenient to
define the $W$ with a normalisation that reflects that of the partition function
(\ref{FK_repr}).

Second, we have to be precise about how the powers of $q$ are computed. In
Figures~\ref{fig:squarekagome}--\ref{fig:hexkagome} we show several examples of
embedded bases $B$. Recall that $G$ is obtained by tiling the two-dimensional
space with copies of $B$. The vertices at the tile boundaries are shared among
two different
copies of $B$; we call those shared vertices the {\em terminals} of
$B$. The embedding can be visualised by pairing the terminals two by
two (shown as matching shapes in
Figures~\ref{fig:squarekagome}--\ref{fig:hexkagome}).
This means that in the embedding a given terminal of one copy of the basis
$B_1$ is identified with the matching terminal of another copy of the
basis $B_2$. In other words, $B_1$ and $B_2$ are glued along matching
terminals. Let now $A$ be a subset of edges of $B$ describing a certain event,
which we classify as 2D, 1D or 0D as above. The weight in the corresponding
$W$ of the event described by $A$ is defined as
\be
 \left( \prod_{i \in A} v_i \right) q^{k(A)-1} \,,
 \label{eq:vq_factors}
\ee
where $k(A)$ is the number of connected components induced by $A$ in the
(generally
non-planar) graph obtained from $B$ by {\em identifying the matching terminals}
as in
the definition of the embedding. Note also that since $k(A) \ge 1$, we have
chosen the
power of $q$ appearing in (\ref{eq:vq_factors}) as $k(A)-1$ rather than $k(A)$.
With this
convention we avoid having an overall factor of $q$ in (\ref{eq:2D0D}).

With the definitions (\ref{eq:2D0D})--(\ref{eq:vq_factors}) the critical polynomial coincides with that defined in \cite{Jacobsen12}.
That the probabilistic and contraction-deletion definitions produce the same polynomial can be seen as follows. First, we note
that the probabilities $W(2D)$ and $W(0D)$ separately satisfy the contraction-deletion identity. This is because these quantities
are just restricted partition functions and will therefore satisfy the same contraction-deletion property as does the full partition function.
In the definition made in \cite{Jacobsen12}, contraction-deletion was used as a recurrence to reduce any basis to a number of three-terminal cases,
for which the exact solvability criterion for 3-uniform hypergraphs was inserted as an initial condition. That latter criterion reads \cite{WuLin80}
$q A - C = 0$, where $A$ and (resp.\ $C$) denotes the weight of all three vertices surrounding a hyper-edge being unconnected
(resp.\ connected). It is easy to see that Eqs.~(\ref{eq:2D0D})--(\ref{eq:vq_factors}) precisely reproduce this initial condition.
Since the recurrence relation (i.e., contraction-deletion) is also identical for the two definitions, it follows that they produce the
same critical polynomial for any choice of the basis $B$.


To see the definitions (\ref{eq:2D0D})--(\ref{eq:vq_factors}) at work, we
consider again 
the 4-edge checkerboard example of Figure~\ref{fig:checkerboard}. We have
\ba
 W(2D;B) &=& v_1 v_2 v_3 v_4 +
 v_2 v_3 v_4 + v_1 v_3 v_4 + v_1 v_2 v_4 + v_1 v_2 v_3 \,, \\
 W(1D;B) &=& v_1 v_2 + v_1 v_3 + v_1 v_4 + v_2 v_3 + v_2 v_4 + v_3 v_4 \,, \\
 W(0D;B) &=& v_1 + v_2 + v_3 + v_4 + q \,.
\ea
Inserting this in (\ref{eq:2D0D}) indeed reproduces (\ref{PB_check}).

Finally we note that for the the special case of percolation, the probabilistic
definition
of $P_B(q,v)$ has already appeared in \cite{SJ12}. The results reported here
reduce
to those of \cite{SJ12} upon setting $q=1$ and $v_i = \frac{p_i}{1-p_i}$, where
$p_i$
is the probability of edge $i$ being open.

\subsection{Bases and embeddings}

As mentioned above, one advantage of the redefinition (\ref{eq:2D0D})
is that we may now use a transfer matrix to compute the critical polynomials
on much larger bases than was possible using contraction-deletion. Below we give
the details of this
approach (section~\ref{sec:tm}) and
report the results for various lattices (section~\ref{sec:results}).

But first we discuss more carefully the bases that we have considered.
We are mainly interested in families of bases whose size can be modulated
by varying one or more integer parameters. This will in particular allow
us to study the size dependence of the resulting $p_c$.

\subsubsection{Square bases}
\label{sec:sq_bases}

\begin{figure}
\begin{center}
\includegraphics{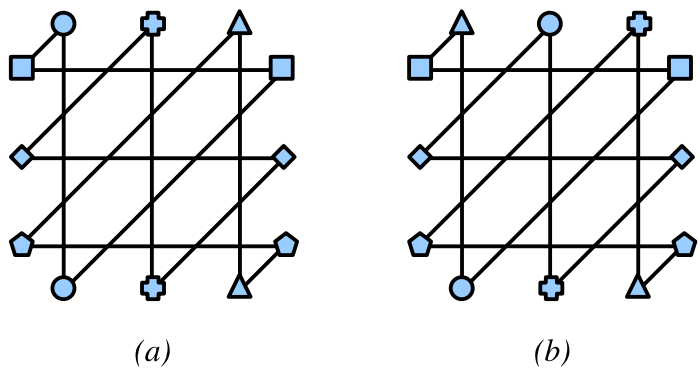}
\caption{$3 \times 3$ square bases for the kagome lattice with: a)
  straight embedding, b) a twisted embedding.}
\label{fig:squarekagome}
\end{center}
\end{figure}

An example of a square basis $B$ is shown in
Figure~\ref{fig:squarekagome}. We recall that the embedding is visualised by
pairing the terminals two by
two (shown as matching shapes in Figure~\ref{fig:squarekagome}).
The embedded basis in Figure~\ref{fig:squarekagome}a is the immediate
generalisation of the checkerboard
example shown in Figure~\ref{fig:checkerboard}, and we refer to
this as the {\em straight embedding}.

A variation of the straight embedding is to shift cyclically the
vertices along one of the sides of the square before gluing them to
those of the opposing side; we call this a {\em twisted embedding}. By
reflection symmetry, shifting cyclically $k$ steps to the right or to
the left produces identical results.  There are thus in general $1 +
\lfloor n/2 \rfloor$ inequivalent twists, corresponding to
$k=0,1,\ldots,\lfloor n/2 \rfloor$. In practice we have found that for
some---but not all---lattices the cases $(n,k) = (2,0)$ and $(n,k) =
(2,1)$ produce the same critical polynomial. But in general the twisting
does change the critical polynomial, as we shall see below.

A square basis $B$ of size $n \times n$ has $n$ terminals on each of
the four sides of the square. The number of vertices and edges in $B$
are both proportional to $n^2$. In the vertex count, each terminal
counts for $1/2$ only, since it is shared among two copies of the
basis. Thus, the square basis for the kagome lattice shown in
Figure~\ref{fig:squarekagome} has $6 n^2$ edges and $3 n^2$ vertices.

One can obviously generalise this construction to rectangular bases of
size $n \times m$. For $n = m$ one recovers a square basis. For $n
\neq m$ the twists along the $n$ and $m$ directions are no longer
equivalent.

\subsubsection{Hexagonal bases}
\label{sec:hex_bases}

\begin{figure}
\begin{center}
\includegraphics{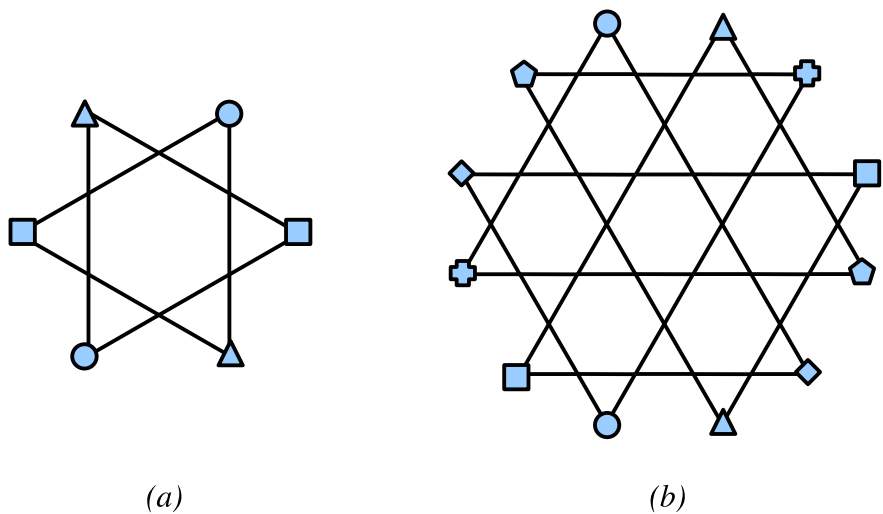}
\caption{Hexagonal bases for the kagome lattice with a) $n=1$, and b) $n=2$.}
\label{fig:hexkagome}
\end{center}
\end{figure}

When the lattice $L$ has a 3-fold rotational symmetry, one can define
as well a hexagonal embedding. Examples of this are shown in
Figure~\ref{fig:hexkagome}. Each of the six sides of the hexagon now
supports $n$ terminals. Note that it is not possible to twist the
hexagonal bases, since only the straight embedding produces a valid
tiling of two-dimensional space.

An advantage of hexagonal bases over the square bases is that they
have a lower ratio of terminals to edges. For example, on the kagome lattice one
has now $6n$
terminals, $9 n^2$ vertices and $18 n^2$ edges. This is useful because the
number of terminals is the limiting factor in the transfer matrix
computation while the accuracy of the critical point estimates increases with
the number of edges.

Another advantage is that the hexagonal basis is designed to respect
the 3-fold rotational symmetry of the lattice.  Thus, for lattices
having this symmetry---such as the kagome and $(3,12^2)$ lattices---we
expect the hexagonal basis to yield better accuracy than the square
basis for a given number of edges. We shall come back to this point
in section~\ref{sec:results}.

Note that one can extend this construction to generalised hexagonal
bases with $2(n_1+n_2+n_3)$ terminals, where each pair of opposing
sides of the hexagon supports $n_i$ terminals $(i=1,2,3)$. The special
case with one of the $n_i = 0$ reproduces the rectangular bases.

\section{Transfer matrix}
\label{sec:tm}

The weights $W(2D;B)$ and $W(0D;B)$ entering the
definition~(\ref{eq:2D0D}) of the critical polynomial can be computed
from a transfer matrix construction along the lines of
Ref.~\cite{BloteNightingale1982}.  First notice that each state of the
edges within the basis $B$ induces a set partition among the
terminals; each part (or block) in the partition consists of a subset
of terminals that are mutually connected through paths of open edges.
The key idea is to first compute the weights of all possible
partitions. One next groups the partitions according to their 2D, 1D
or 0D nature in order to evaluate (\ref{eq:2D0D}).

With $N$ terminals, the number of partitions respecting planarity is
given by the Catalan number
\begin{equation}
 C_N = \frac{1}{N+1} {2N \choose N} \,.
\end{equation}
For example, the $C_3 = 5$ planar partitions of the set $\{1,2,3\}$
are denoted
\begin{equation}
 (1)(2)(3) \,, \quad (12)(3) \,, \quad (13)(2) \,, \quad (1)(23) \,, \quad (123)
\,,
\end{equation}
where the elements belonging to the same part are grouped inside
parentheses.

The dimension of the transfer matrix is thus $C_N$, and both time and
memory requirements are proportional to this number.%
\footnote{We assume here the use of standard sparse matrix
  factorisation techniques \cite{JacobsenCardy1998}.}
Asymptotically we have $C_N \sim 4^N$ for $N \gg 1$. Taking as an
example the kagome lattice with the $n \times n$ square basis, the
time complexity of the transfer matrix method is then $\sim 4^{4n} =
2^{8n}$. This can be compared to the contraction-deletion method,
whose number of recursive calls is $\sim 2^{6 n^2}$.

\subsection{Square bases}

\begin{figure}
\begin{center}
 \begin{tikzpicture}[scale=0.5]
  \foreach \xpos in {0,2}
  \foreach \ypos in {0,3.464}
  {
   \draw[fill,blue!20,line width=0ex]
(\xpos-0.5,\ypos+0.866)--(\xpos+0.5,\ypos+2.598)--(\xpos+1.5,
\ypos+0.866)--(\xpos+0.5,\ypos-0.866)--cycle;
   \draw[blue,dashed,line width=0.2ex]
(\xpos-0.5,\ypos+0.866)--(\xpos+0.5,\ypos+2.598)--(\xpos+1.5,
\ypos+0.866)--(\xpos+0.5,\ypos-0.866)--cycle;
   \draw[black,line width=0.3ex]
(\xpos,\ypos)--(\xpos+1,\ypos)--(\xpos,\ypos+1.732)--(\xpos+1,
\ypos+1.732)--cycle;
  }
  \foreach \xpos in {-1,1,3}
  \foreach \ypos in {1.732}
  {
   \draw[fill,red!10,line width=0ex]
(\xpos-0.5,\ypos+0.866)--(\xpos+0.5,\ypos+2.598)--(\xpos+1.5,
\ypos+0.866)--(\xpos+0.5,\ypos-0.866)--cycle;
   \draw[blue,dashed,line width=0.2ex]
(\xpos-0.5,\ypos+0.866)--(\xpos+0.5,\ypos+2.598)--(\xpos+1.5,
\ypos+0.866)--(\xpos+0.5,\ypos-0.866)--cycle;
   \draw[black,line width=0.3ex]
(\xpos,\ypos)--(\xpos+1,\ypos)--(\xpos,\ypos+1.732)--(\xpos+1,
\ypos+1.732)--cycle;
  }
  \foreach \xpos in {1}
  \foreach \ypos in {-1.732,5.196}
  {
   \draw[fill,red!10,line width=0ex]
(\xpos-0.5,\ypos+0.866)--(\xpos+0.5,\ypos+2.598)--(\xpos+1.5,
\ypos+0.866)--(\xpos+0.5,\ypos-0.866)--cycle;
   \draw[blue,dashed,line width=0.2ex]
(\xpos-0.5,\ypos+0.866)--(\xpos+0.5,\ypos+2.598)--(\xpos+1.5,
\ypos+0.866)--(\xpos+0.5,\ypos-0.866)--cycle;
   \draw[black,line width=0.3ex]
(\xpos,\ypos)--(\xpos+1,\ypos)--(\xpos,\ypos+1.732)--(\xpos+1,
\ypos+1.732)--cycle;
  }
 \draw (-1,1.732) node[below left] {$1$};
 \draw (0,0) node[below left] {$2$};
 \draw (1,-1.732) node[below left] {$3$};
 \draw (2,-1.732) node[below right] {$4$};
 \draw (3,0) node[below right] {$5$};
 \draw (4,1.732) node[below right] {$6$};
 \draw (-1,3.464) node[above left] {$1'$};
 \draw (0,5.196) node[above left] {$2'$};
 \draw (1,6.928) node[above left] {$3'$};
 \draw (2,6.928) node[above right] {$4'$};
 \draw (3,5.196) node[above right] {$5'$};
 \draw (4,3.464) node[above right] {$6'$};
 \begin{scope}[xshift=12cm,yshift=1.732cm]
  \draw (-0.5,0.866) node[left] {${\sf B}_i =$};
  \draw (0,0) node[below left] {$i$};
  \draw (1,0) node[below right] {$i+1$};
  \draw (0,1.732) node[above left] {$i'$};
  \draw (1,1.732) node[above right] {$i'+1$};
  \foreach \xpos in {0}
  \foreach \ypos in {0}
  {
   \draw[fill,red!0,line width=0ex]
(\xpos-0.5,\ypos+0.866)--(\xpos+0.5,\ypos+2.598)--(\xpos+1.5,
\ypos+0.866)--(\xpos+0.5,\ypos-0.866)--cycle;
   \draw[blue,dashed,line width=0.2ex]
(\xpos-0.5,\ypos+0.866)--(\xpos+0.5,\ypos+2.598)--(\xpos+1.5,
\ypos+0.866)--(\xpos+0.5,\ypos-0.866)--cycle;
   \draw[black,line width=0.3ex]
(\xpos,\ypos)--(\xpos+1,\ypos)--(\xpos,\ypos+1.732)--(\xpos+1,
\ypos+1.732)--cycle;
  }
 \end{scope}
 \end{tikzpicture}
\end{center}
 \caption{Transfer matrix construction for the kagome lattice on
   an $n \times n$ square basis, here with $n=3$. The operator ${\sf
     B}_i$ adds six edges to the lattice.}
 \label{fig:kagomeTM}
\end{figure}
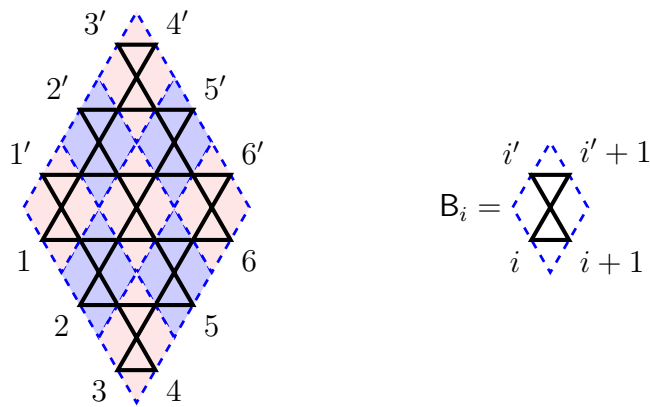

Our transfer matrix construction is most easily explained on a specific
example. So consider the kagome lattice with the $n \times n$ square basis;
the case $n=3$ is shown in Figure~\ref{fig:kagomeTM}.

The transfer matrix ${\sf T}$ constructs the lattice from the bottom
to the top, while keeping track of the Boltzmann weight of each
partition of the terminals.  The bottom terminals are denoted
$1,2,\ldots,2n$ and the top terminals $1',2',\ldots,2n'$. At the
beginning of the process the top and bottom are identified, so the
initial state $|{\rm i} \rangle$ on which ${\sf T}$ acts is the partition
$(1\,1')(2\,2')\cdots(2n\,2n')$ with weight $1$.

We now define two kinds of operators acting on a partition
\cite{SalasSokal2001}:
\begin{itemize}
 \item The join operator ${\sf J}_i$ amalgamates the parts to which
   the top terminals $i'$ and $i'+1$ belong. In particular, on
   partitions in which those two terminals already belong to the same part,
   ${\sf J}_i$ acts as the identity operator ${\sf I}$. Note that if some parts
   contain both bottom and top terminals, the action of ${\sf J}_i$
   can also affect the connections among the bottom terminals.
 \item The detach operator ${\sf D}_i$ detaches the top terminal $i'$
   from its part and transforms it into a singleton in the
   partition. But if that terminal was already a singleton,
   ${\sf D}_i$ acts as $q \, {\sf I}$. The reason for the factor of $q$ is that
a detached singleton amounts to
   a connected component being ``seen for the last time'', and ${\sf D}_i$
   should then apply the corresponding weight appearing in
(\ref{eq:vq_factors}).
\end{itemize}
{}From these two basic operators and the identity operator ${\sf I}$
we now define an operator
\begin{equation}
 {\sf H}_i = {\sf I} + v {\sf J}_i
 \label{eq:H_op}
\end{equation}
that adds a horizontal edge to the lattice. The word ``horizontal''
refers to a drawing of the lattice where the top terminals $i'$ and
$i'+1$ are horizontally aligned; otherwise the edge would be
better described as ``diagonal''.  Note that ${\sf H}_i$ attaches a
weight $1$ (resp.\ $v$) to a closed (resp.\ open) horizontal edge,
as required. Similarly we define
\begin{equation}
 {\sf V}_i = v {\sf I} + {\sf D}_i
 \label{eq:V_op}
\end{equation}
that adds a vertical edge between $i'$ and $i''$, where $i'$
(resp.\ $i''$) denotes the corresponding top terminal before
(resp.\ after) the action of ${\sf V}_i$. To simplify the notation, it
is convenient to assume that following the action of ${\sf V}_i$ we
relabel $i''$ as $i'$. The word ``vertical'' refers to a drawing of the
lattice where $i'$ and $i''$ are vertically aligned.

The fundamental building block of the lattice shown on the right of
Figure~\ref{fig:kagomeTM} is then constructed by the composite
operator
\begin{equation}
 {\sf B}_i = {\sf H}_i {\sf V}_i {\sf H}_i {\sf D}_{i+1} {\sf H}_i {\sf V}_i
{\sf H}_i \,,
 \qquad \mbox{Kagome lattice}
\end{equation}
where the operators here and elsewhere should be understood as acting in order
from right to left. The whole lattice $B$ is finally obtained by adding
successive rows
(for clarity shown in alternating hues on the left of
Figure~\ref{fig:kagomeTM}) of ${\sf B}_i$. The transfer matrix then reads%
\footnote{To avoid any ambiguity about the ordering of operators we write out
the rightmost double product in (\ref{TM_order}):
${\sf B}_1 {\sf B}_3 {\sf B}_5 \cdots {\sf B}_{2n-1} \times  \cdots \times {\sf
B}_{n-1} {\sf B}_{n+1} \times {\sf B}_n$. This should
be compared with Figure~\ref{fig:kagomeTM}, and we recall that the rightmost
factor acts first.}
\begin{equation}
 {\sf T} = \prod_{y=1}^{n-1} \prod_{x=1}^{y} {\sf B}_{n-y-1+2x} \times
           \prod_{y=1}^n \prod_{x=0}^{n-y} {\sf B}_{y+2x}
           \label{TM_order}
\end{equation}
and the final state
\begin{equation}
 |{\rm f}\rangle = {\sf T} |{\rm i}\rangle
 \label{final_state}
\end{equation}
contains all possible partitions among the $4n$ terminals along with their
respective Boltzmann weights.

The final state $|{\rm f}\rangle$ contains all the information necessary to
extract $P_B(q,v)$,
and the remainder of section~\ref{sec:tm} explains how this is done. But first
we describe how to
modify the transfer matrix formalism just described to accommodate other
lattices
(section~\ref{sec:other_lattices}) and hexagonal bases
(section~\ref{sec:TM_hex_bases}).
Then, in section~\ref{sec:2D1D0D}, we show how to determine which of the states
in
$|{\rm f}\rangle$ contribute to the weights $W(2D;B)$ and $W(0D;B)$ appearing in
(\ref{eq:2D0D}).

Finally, each contributing state needs to be counted with the correct weight
(\ref{eq:vq_factors}).
The powers of $v$ are unproblematic and have been explicitly accounted for in
(\ref{eq:H_op})--(\ref{eq:V_op}).
The contributions to $q^{k(A)-1}$ originating from connected components not
containing any terminal
of $B$ have been accounted for in the above definition of ${\sf D}_i$. It
remains to explain how to count
the connected components containing at least one terminal; this is the subject
of section~\ref{sec:q_power}.

\subsubsection{Other lattices}
\label{sec:other_lattices}

The extension of the transfer matrix formalism to the other lattices considered
in this paper is very
simple: it suffices to change the definition of the operator ${\sf
  B}_i$, while leaving the remainder of the construction unchanged.%
\footnote{In practice, when implementing this algorithm on a computer,
  this implies that only a few lines of code have to be modified to
  change the lattice.}

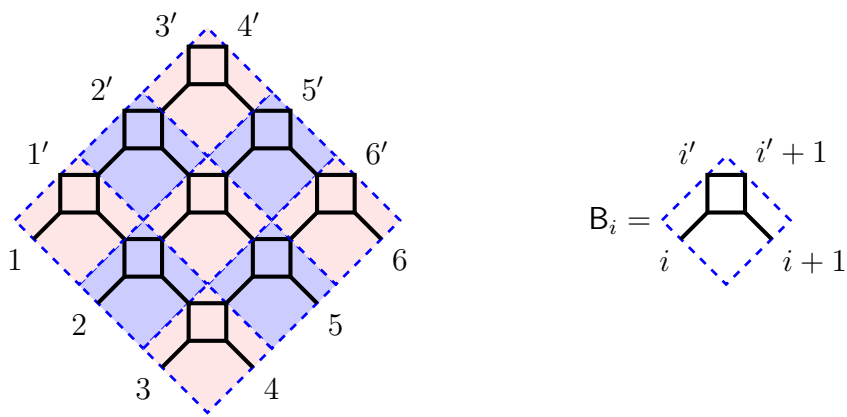
\begin{figure}
\begin{center}
 \begin{tikzpicture}[scale=0.5]
  \foreach \xpos in {0,3.414}
  \foreach \ypos in {0,3.414}
  {
   \draw[fill,blue!20,line width=0ex]
(\xpos-0.5,\ypos+0.5)--(\xpos+1.207,\ypos+2.207)--(\xpos+2.914,
\ypos+0.5)--(\xpos+1.207,\ypos-1.207)--cycle;
   \draw[blue,dashed,line width=0.2ex]
(\xpos-0.5,\ypos+0.5)--(\xpos+1.207,\ypos+2.207)--(\xpos+2.914,
\ypos+0.5)--(\xpos+1.207,\ypos-1.207)--cycle;
   \draw[black,line width=0.3ex]
(\xpos,\ypos)--(\xpos+0.707,\ypos+0.707)--(\xpos+1.707,
\ypos+0.707)--(\xpos+2.414,\ypos);
   \draw[black,line width=0.3ex]
(\xpos+0.707,\ypos+0.707)--(\xpos+0.707,\ypos+1.707)--(\xpos+1.707,
\ypos+1.707)--(\xpos+1.707,\ypos+0.707);
  }
  \foreach \xpos in {-1.707,1.707,5.121}
  \foreach \ypos in {1.707}
  {
   \draw[fill,red!10,line width=0ex]
(\xpos-0.5,\ypos+0.5)--(\xpos+1.207,\ypos+2.207)--(\xpos+2.914,
\ypos+0.5)--(\xpos+1.207,\ypos-1.207)--cycle;
   \draw[blue,dashed,line width=0.2ex]
(\xpos-0.5,\ypos+0.5)--(\xpos+1.207,\ypos+2.207)--(\xpos+2.914,
\ypos+0.5)--(\xpos+1.207,\ypos-1.207)--cycle;
   \draw[black,line width=0.3ex]
(\xpos,\ypos)--(\xpos+0.707,\ypos+0.707)--(\xpos+1.707,
\ypos+0.707)--(\xpos+2.414,\ypos);
   \draw[black,line width=0.3ex]
(\xpos+0.707,\ypos+0.707)--(\xpos+0.707,\ypos+1.707)--(\xpos+1.707,
\ypos+1.707)--(\xpos+1.707,\ypos+0.707);
  }
  \foreach \xpos in {1.707}
  \foreach \ypos in {-1.707,5.121}
  {
   \draw[fill,red!10,line width=0ex]
(\xpos-0.5,\ypos+0.5)--(\xpos+1.207,\ypos+2.207)--(\xpos+2.914,
\ypos+0.5)--(\xpos+1.207,\ypos-1.207)--cycle;
   \draw[blue,dashed,line width=0.2ex]
(\xpos-0.5,\ypos+0.5)--(\xpos+1.207,\ypos+2.207)--(\xpos+2.914,
\ypos+0.5)--(\xpos+1.207,\ypos-1.207)--cycle;
   \draw[black,line width=0.3ex]
(\xpos,\ypos)--(\xpos+0.707,\ypos+0.707)--(\xpos+1.707,
\ypos+0.707)--(\xpos+2.414,\ypos);
   \draw[black,line width=0.3ex]
(\xpos+0.707,\ypos+0.707)--(\xpos+0.707,\ypos+1.707)--(\xpos+1.707,
\ypos+1.707)--(\xpos+1.707,\ypos+0.707);
  }
 \draw (-1.707,1.707) node[below left] {$1$};
 \draw (0,0) node[below left] {$2$};
 \draw (1.707,-1.707) node[below left] {$3$};
 \draw (4.121,-1.707) node[below right] {$4$};
 \draw (5.828,0) node[below right] {$5$};
 \draw (7.535,1.707) node[below right] {$6$};
 \draw (-1.000,3.414) node[above left] {$1'$};
 \draw (0.707,5.121) node[above left] {$2'$};
 \draw (2.414,6.828) node[above left] {$3'$};
 \draw (3.414,6.828) node[above right] {$4'$};
 \draw (5.121,5.121) node[above right] {$5'$};
 \draw (6.828,3.414) node[above right] {$6'$};
  \begin{scope}[xshift=15.5cm,yshift=1.707cm]
  \draw (-0.5,0.5) node[left] {${\sf B}_i =$};
  \draw (0,0) node[below left] {$i$};
  \draw (2.414,0) node[below right] {$i+1$};
  \draw (0.707,1.707) node[above left] {$i'$};
  \draw (1.707,1.707) node[above right] {$i'+1$};
  \foreach \xpos in {0}
  \foreach \ypos in {0}
  {
   \draw[fill,red!0,line width=0ex]
(\xpos-0.5,\ypos+0.5)--(\xpos+1.207,\ypos+2.207)--(\xpos+2.914,
\ypos+0.5)--(\xpos+1.207,\ypos-1.207)--cycle;
   \draw[blue,dashed,line width=0.2ex]
(\xpos-0.5,\ypos+0.5)--(\xpos+1.207,\ypos+2.207)--(\xpos+2.914,
\ypos+0.5)--(\xpos+1.207,\ypos-1.207)--cycle;
   \draw[black,line width=0.3ex]
(\xpos,\ypos)--(\xpos+0.707,\ypos+0.707)--(\xpos+1.707,
\ypos+0.707)--(\xpos+2.414,\ypos);
   \draw[black,line width=0.3ex]
(\xpos+0.707,\ypos+0.707)--(\xpos+0.707,\ypos+1.707)--(\xpos+1.707,
\ypos+1.707)--(\xpos+1.707,\ypos+0.707);
  }
 \end{scope}
 \end{tikzpicture}
\end{center}
 \caption{Transfer matrix construction for the $(4,8^2)$ lattice on
   an $n \times n$ square basis, here with $n=3$. The operator ${\sf
     B}_i$ adds six edges to the lattice.}
 \label{fig:foureightTM}
\end{figure}

The square basis for the $(4,8^2)$ lattice is shown in
Figure~\ref{fig:foureightTM}.
Its fundamental building block now has the expression
\begin{equation}
 {\sf B}_i = {\sf H}_i {\sf V}_i {\sf V}_{i+1} {\sf H}_i {\sf V}_i {\sf V}_{i+1}
\,,
 \qquad \mbox{$(4,8^2)$ lattice.}
\end{equation}

\begin{figure}
\begin{center}
 \begin{tikzpicture}[scale=0.5]
  \foreach \xpos in {0,3.732}
  \foreach \ypos in {0,6.464}
  {
   \draw[fill,blue!20,line width=0ex]
(\xpos-0.5,\ypos+0.866)--(\xpos+1.366,\ypos-2.366)--(\xpos+3.232,
\ypos+0.866)--(\xpos+1.366,\ypos+4.098)--cycle;
   \draw[blue,dashed,line width=0.2ex]
(\xpos-0.5,\ypos+0.866)--(\xpos+1.366,\ypos-2.366)--(\xpos+3.232,
\ypos+0.866)--(\xpos+1.366,\ypos+4.098)--cycle;
   \draw[black,line width=0.3ex]
(\xpos,\ypos)--(\xpos+0.866,\ypos+0.5)--(\xpos+1.866,\ypos+0.5)--(\xpos+1.366,
\ypos+1.366)--(\xpos+0.866,\ypos+0.5);
   \draw[black,line width=0.3ex] (\xpos+1.866,\ypos+0.5)--(\xpos+2.732,\ypos);
   \draw[black,line width=0.3ex]
(\xpos+1.366,\ypos+1.366)--(\xpos+1.366,\ypos+2.366)--(\xpos+1.866,
\ypos+3.232)--(\xpos+0.866,\ypos+3.232)--(\xpos+1.366,\ypos+2.366);
  }
  \foreach \xpos in {-1.866,1.866,5.598}
  \foreach \ypos in {3.232}
  {
   \draw[fill,red!10,line width=0ex]
(\xpos-0.5,\ypos+0.866)--(\xpos+1.366,\ypos-2.366)--(\xpos+3.232,
\ypos+0.866)--(\xpos+1.366,\ypos+4.098)--cycle;
   \draw[blue,dashed,line width=0.2ex]
(\xpos-0.5,\ypos+0.866)--(\xpos+1.366,\ypos-2.366)--(\xpos+3.232,
\ypos+0.866)--(\xpos+1.366,\ypos+4.098)--cycle;
   \draw[black,line width=0.3ex]
(\xpos,\ypos)--(\xpos+0.866,\ypos+0.5)--(\xpos+1.866,\ypos+0.5)--(\xpos+1.366,
\ypos+1.366)--(\xpos+0.866,\ypos+0.5);
   \draw[black,line width=0.3ex] (\xpos+1.866,\ypos+0.5)--(\xpos+2.732,\ypos);
   \draw[black,line width=0.3ex]
(\xpos+1.366,\ypos+1.366)--(\xpos+1.366,\ypos+2.366)--(\xpos+1.866,
\ypos+3.232)--(\xpos+0.866,\ypos+3.232)--(\xpos+1.366,\ypos+2.366);
  }
  \foreach \xpos in {1.866}
  \foreach \ypos in {-3.232,9.696}
  {
   \draw[fill,red!10,line width=0ex]
(\xpos-0.5,\ypos+0.866)--(\xpos+1.366,\ypos-2.366)--(\xpos+3.232,
\ypos+0.866)--(\xpos+1.366,\ypos+4.098)--cycle;
   \draw[blue,dashed,line width=0.2ex]
(\xpos-0.5,\ypos+0.866)--(\xpos+1.366,\ypos-2.366)--(\xpos+3.232,
\ypos+0.866)--(\xpos+1.366,\ypos+4.098)--cycle;
   \draw[black,line width=0.3ex]
(\xpos,\ypos)--(\xpos+0.866,\ypos+0.5)--(\xpos+1.866,\ypos+0.5)--(\xpos+1.366,
\ypos+1.366)--(\xpos+0.866,\ypos+0.5);
   \draw[black,line width=0.3ex] (\xpos+1.866,\ypos+0.5)--(\xpos+2.732,\ypos);
   \draw[black,line width=0.3ex]
(\xpos+1.366,\ypos+1.366)--(\xpos+1.366,\ypos+2.366)--(\xpos+1.866,
\ypos+3.232)--(\xpos+0.866,\ypos+3.232)--(\xpos+1.366,\ypos+2.366);
  }
 \draw (-1.866,3.232) node[below left] {$1$};
 \draw (0,0) node[below left] {$2$};
 \draw (1.866,-3.232) node[below left] {$3$};
 \draw (4.598,-3.232) node[below right] {$4$};
 \draw (6.464,0) node[below right] {$5$};
 \draw (8.330,3.232) node[below right] {$6$};
 \draw (-1.000,6.464) node[above left] {$1'$};
 \draw (0.866,9.696) node[above left] {$2'$};
 \draw (2.732,12.928) node[above left] {$3'$};
 \draw (3.732,12.928) node[above right] {$4'$};
 \draw (5.598,9.696) node[above right] {$5'$};
 \draw (7.464,6.464) node[above right] {$6'$};
  \begin{scope}[xshift=16.5cm,yshift=3.232cm]
  \draw (-0.5,0.866) node[left] {${\sf B}_i =$};
  \draw (0,0) node[below left] {$i$};
  \draw (2.732,0) node[below right] {$i+1$};
  \draw (0.866,3.232) node[above left] {$i'$};
  \draw (1.866,3.232) node[above right] {$i'+1$};
  \foreach \xpos in {0}
  \foreach \ypos in {0}
  {
   \draw[fill,red!0,line width=0ex]
(\xpos-0.5,\ypos+0.866)--(\xpos+1.366,\ypos-2.366)--(\xpos+3.232,
\ypos+0.866)--(\xpos+1.366,\ypos+4.098)--cycle;
   \draw[blue,dashed,line width=0.2ex]
(\xpos-0.5,\ypos+0.866)--(\xpos+1.366,\ypos-2.366)--(\xpos+3.232,
\ypos+0.866)--(\xpos+1.366,\ypos+4.098)--cycle;
   \draw[black,line width=0.3ex]
(\xpos,\ypos)--(\xpos+0.866,\ypos+0.5)--(\xpos+1.866,\ypos+0.5)--(\xpos+1.366,
\ypos+1.366)--(\xpos+0.866,\ypos+0.5);
   \draw[black,line width=0.3ex] (\xpos+1.866,\ypos+0.5)--(\xpos+2.732,\ypos);
   \draw[black,line width=0.3ex]
(\xpos+1.366,\ypos+1.366)--(\xpos+1.366,\ypos+2.366)--(\xpos+1.866,
\ypos+3.232)--(\xpos+0.866,\ypos+3.232)--(\xpos+1.366,\ypos+2.366);
  }
 \end{scope}
 \end{tikzpicture}
\end{center}
 \caption{Transfer matrix construction for the $(3,12^2)$ lattice on
   an $n \times n$ square basis, here with $n=3$. The operator ${\sf
     B}_i$ adds nine edges to the lattice.}
 \label{fig:threetwelveTM}
\end{figure}
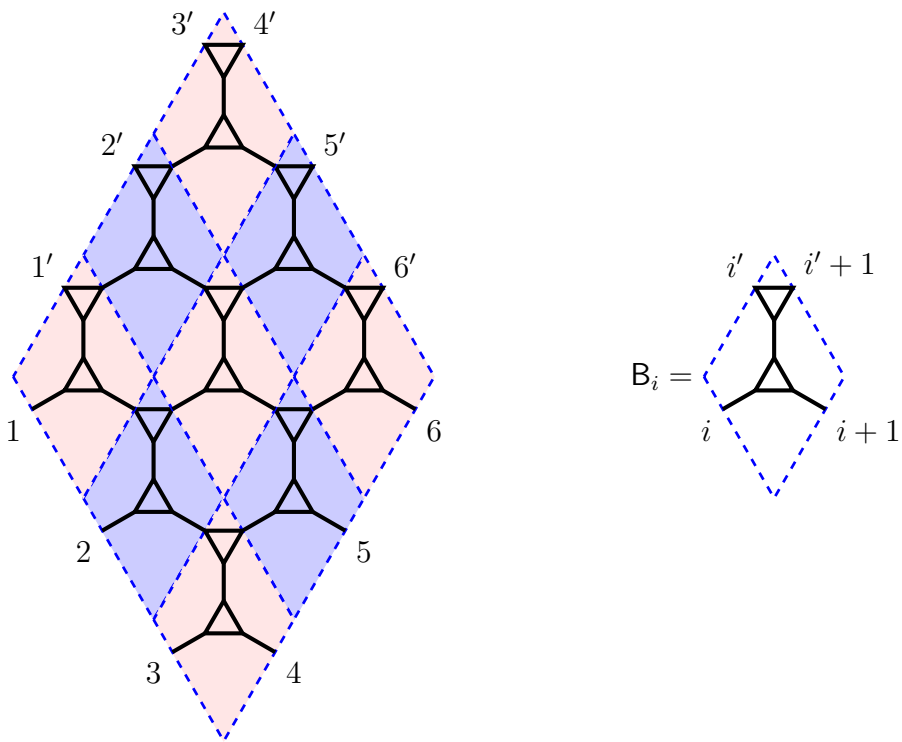

As a last example, consider the $(3,12^2)$ lattice with the square basis
depicted
in Figure~\ref{fig:threetwelveTM}. We find in this case
\begin{equation}
 {\sf B}_i = {\sf H}_i {\sf V}_i {\sf H}_i {\sf V}_i {\sf D}_{i+1}
             {\sf H}_i {\sf V}_i {\sf H}_i {\sf V}_i {\sf V}_{i+1} \,,
 \qquad \mbox{$(3,12^2)$ lattice.}
\end{equation}

\subsection{Hexagonal bases}
\label{sec:TM_hex_bases}

Because of their 3-fold rotational symmetry, it is also interesting to study the
kagome and $(3,12^2)$ lattice with a hexagonal basis. We now describe how to
adapt the transfer matrix construction to this case.

\begin{figure}
\begin{center}
 \begin{tikzpicture}[scale=0.5]
  \foreach \xpos in {-3.464,0,3.464}
  \foreach \ypos in {2,4,6,8}
  {
   \draw[fill,blue!20,line width=0ex]
(\xpos,\ypos)--(\xpos+1.732,\ypos+1)--(\xpos+3.464,\ypos)--(\xpos+1.732,
\ypos-1)--cycle;
   \draw[blue,dashed,line width=0.2ex]
(\xpos,\ypos)--(\xpos+1.732,\ypos+1)--(\xpos+3.464,\ypos)--(\xpos+1.732,
\ypos-1)--cycle;
   \draw[black,line width=0.3ex]
(\xpos+0.866,\ypos-0.5)--(\xpos+2.598,\ypos+0.5)--(\xpos+2.598,
\ypos-0.5)--(\xpos+0.866,\ypos+0.5)--cycle;
  }
  \foreach \xpos in {-1.732,1.732}
  \foreach \ypos in {1,3,5,7,9}
  {
   \draw[fill,red!10,line width=0ex]
(\xpos,\ypos)--(\xpos+1.732,\ypos+1)--(\xpos+3.464,\ypos)--(\xpos+1.732,
\ypos-1)--cycle;
   \draw[blue,dashed,line width=0.2ex]
(\xpos,\ypos)--(\xpos+1.732,\ypos+1)--(\xpos+3.464,\ypos)--(\xpos+1.732,
\ypos-1)--cycle;
   \draw[black,line width=0.3ex]
(\xpos+0.866,\ypos-0.5)--(\xpos+2.598,\ypos+0.5)--(\xpos+2.598,
\ypos-0.5)--(\xpos+0.866,\ypos+0.5)--cycle;
  }
  \foreach \xpos in {0}
  \foreach \ypos in {0,10}
  {
   \draw[fill,blue!20,line width=0ex]
(\xpos,\ypos)--(\xpos+1.732,\ypos+1)--(\xpos+3.464,\ypos)--(\xpos+1.732,
\ypos-1)--cycle;
   \draw[blue,dashed,line width=0.2ex]
(\xpos,\ypos)--(\xpos+1.732,\ypos+1)--(\xpos+3.464,\ypos)--(\xpos+1.732,
\ypos-1)--cycle;
   \draw[black,line width=0.3ex]
(\xpos+0.866,\ypos-0.5)--(\xpos+2.598,\ypos+0.5)--(\xpos+2.598,
\ypos-0.5)--(\xpos+0.866,\ypos+0.5)--cycle;
  }
  \foreach \xpos in {-3.464}
  \foreach \ypos in {3,5,7}
  {
   \draw[fill,red!10,line width=0ex]
(\xpos,\ypos-1)--(\xpos,\ypos+1)--(\xpos+1.732,\ypos)--cycle;
   \draw[blue,dashed,line width=0.2ex]
(\xpos,\ypos-1)--(\xpos,\ypos+1)--(\xpos+1.732,\ypos)--cycle;
   \draw[black,line width=0.3ex]
(\xpos,\ypos)--(\xpos+0.866,\ypos+0.5)--(\xpos+0.866,\ypos-0.5)--cycle;
  }
  \foreach \xpos in {6.928}
  \foreach \ypos in {3,5,7}
  {
   \draw[fill,red!10,line width=0ex]
(\xpos,\ypos-1)--(\xpos,\ypos+1)--(\xpos-1.732,\ypos)--cycle;
   \draw[blue,dashed,line width=0.2ex]
(\xpos,\ypos-1)--(\xpos,\ypos+1)--(\xpos-1.732,\ypos)--cycle;
   \draw[black,line width=0.3ex]
(\xpos,\ypos)--(\xpos-0.866,\ypos+0.5)--(\xpos-0.866,\ypos-0.5)--cycle;
  }
 \begin{scope}[xshift=14cm,yshift=9cm]
  \draw (0,0) node[left] {${\sf B}_i =$};
  \draw (0.866,-0.5) node[below left] {$i$};
  \draw (2.598,-0.5) node[below right] {$i+1$};
  \draw (0.866,0.5) node[above left] {$i'$};
  \draw (2.598,0.5) node[above right] {$i'+1$};
  \foreach \xpos in {0}
  \foreach \ypos in {0}
  {
   \draw[fill,blue!0,line width=0ex]
(\xpos,\ypos)--(\xpos+1.732,\ypos+1)--(\xpos+3.464,\ypos)--(\xpos+1.732,
\ypos-1)--cycle;
   \draw[blue,dashed,line width=0.2ex]
(\xpos,\ypos)--(\xpos+1.732,\ypos+1)--(\xpos+3.464,\ypos)--(\xpos+1.732,
\ypos-1)--cycle;
   \draw[black,line width=0.3ex]
(\xpos+0.866,\ypos-0.5)--(\xpos+2.598,\ypos+0.5)--(\xpos+2.598,
\ypos-0.5)--(\xpos+0.866,\ypos+0.5)--cycle;
  }
 \end{scope}
 \begin{scope}[xshift=14cm,yshift=5cm]
  \draw (0,0) node[left] {${\sf L}_i =$};
  \draw (1.732,0) node[left] {$i$};
  \draw (2.598,-0.5) node[below right] {$i+1$};
  \draw (2.598,0.5) node[above right] {$i'+1$};
  \foreach \xpos in {1.732}
  \foreach \ypos in {0}
  {
   \draw[fill,red!0,line width=0ex]
(\xpos,\ypos-1)--(\xpos,\ypos+1)--(\xpos+1.732,\ypos)--cycle;
   \draw[blue,dashed,line width=0.2ex]
(\xpos,\ypos-1)--(\xpos,\ypos+1)--(\xpos+1.732,\ypos)--cycle;
   \draw[black,line width=0.3ex]
(\xpos,\ypos)--(\xpos+0.866,\ypos+0.5)--(\xpos+0.866,\ypos-0.5)--cycle;
  }
 \end{scope}
 \begin{scope}[xshift=14cm,yshift=1cm]
  \draw (0,0) node[left] {${\sf R}_i =$};
  \draw (1.732,0) node[right] {$i+1$};
  \draw (0.866,-0.5) node[below left] {$i$};
  \draw (0.866,0.5) node[above left] {$i'$};
  \foreach \xpos in {1.732}
  \foreach \ypos in {0}
  {
   \draw[fill,red!0,line width=0ex]
(\xpos,\ypos-1)--(\xpos,\ypos+1)--(\xpos-1.732,\ypos)--cycle;
   \draw[blue,dashed,line width=0.2ex]
(\xpos,\ypos-1)--(\xpos,\ypos+1)--(\xpos-1.732,\ypos)--cycle;
   \draw[black,line width=0.3ex]
(\xpos,\ypos)--(\xpos-0.866,\ypos+0.5)--(\xpos-0.866,\ypos-0.5)--cycle;
  }
 \end{scope}
 \end{tikzpicture}
\end{center}
\caption{Kagome lattice on a hexagonal basis of size $n$, here with $n=3$. The
operator
${\sf B}_i$ adds six edges to the lattice, while the left and right boundary
operators,
${\sf L}_i$ and ${\sf R}_i$, each add three.}
\label{fig:kagome_hex}
\end{figure}
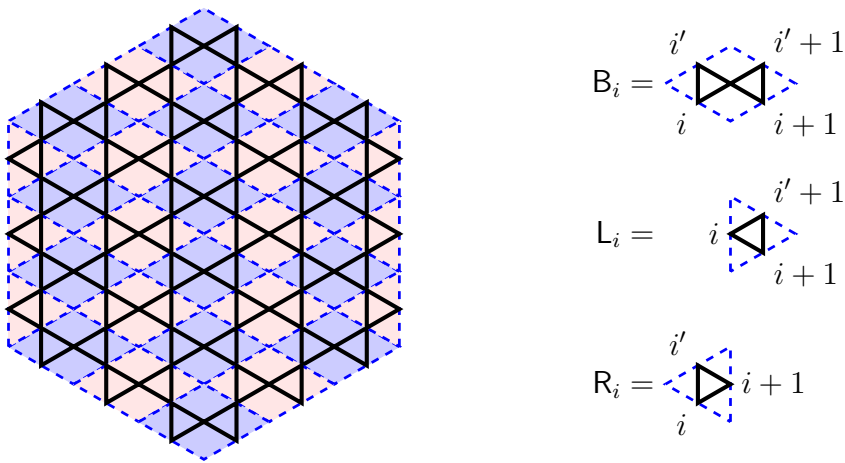

Consider as an example the kagome lattice with the hexagonal basis of
size $n$; the case $n=3$ is shown in
Figure~\ref{fig:kagome_hex}. There are now $6n$ terminals.  Those on
the two bottom sides (resp.\ the two top sides) of the hexagon are
labelled $1,2,\ldots,2n$ (resp.\ $1',2',\ldots,2n'$), just as in the
case of the square basis. We describe below how the remaining
terminals on the left and right sides of the hexagon are to be
handled. The transfer matrix ${\sf T}$ still constructs the lattice
from the bottom to the top.

The expression for the building block ${\sf B}_i$ now needs some
modification, since the orientation of the bow tie motif with respect
to the transfer direction (invariably upwards) has been changed. One
easy option would be to handle the centre of the bow tie as an extra
point---we would then label the three points $i$, $i+1$ and
$i+2$)---and use the expression ${\sf B}_i = {\sf D}_{i+1} {\sf
  H}_{i+1} {\sf H}_i {\sf V}_{i+2} {\sf V}_i {\sf H}_{i+1} {\sf H}_i$.
It is however more efficient to avoid introducing the centre point
into the partition (and keep the usual labelling $i$, $i+1$ as shown
on the right of Figure~\ref{fig:kagome_hex}). The expression for ${\sf
  B}_i$ can then be found by computing the final state
(\ref{final_state}) for the $1 \times 1$ square basis
and rotating the labels (we denote here $j = i+1$):
\begin{eqnarray}
 {\sf B}_i &=& (v^6 + 6 v^5 + 9 v^4) (iji'j') + (2v^4+6v^3+q v^2) (ii')(jj')
\nonumber \\
           &+& (v^4 + 3 v^3) \big[ (i)(ji'j') +
               (j)(ii'j') + (i')(ijj') + (j')(iji') \big] \nonumber \\
           &+& (v^3 + 5 v^2 + q v) \big[ (ii')(j)(j') + (i)(i')(jj') \big] +
               (4v+q) (i)(j)(i')(j') \nonumber \\
           &+& v^2 \left[ (i)(j)(i'j') + (ij)(i')(j') + (i)(i'j)(j') +
(i')(ij')(j) \right] \,.
 \label{newBkagome}
\end{eqnarray}
where a bracketed operator, for example $(ii')(jj')$, creates a bow-tie between
$i$ and $j$ with the indicated partition of its four bounding vertices.
On the boundary of the hexagon we need the further operators
\begin{eqnarray}
 {\sf L}_i &=& {\sf H}_i {\sf V}_{i+1} {\sf H}_i \,, \\
 {\sf R}_i &=& {\sf H}_i {\sf V}_i {\sf H}_i \,.
\end{eqnarray}
The transfer matrix that builds the whole hexagon then reads
\begin{eqnarray}
 {\sf T} &=& \prod_{y=1}^{n-1} \prod_{x=1}^{y} {\sf B}_{n-y-1+2x} \nonumber \\
    &\times& \prod_{y=1}^n \left( \prod_{x=1}^n {\sf B}_{2x-1} \times
             {\sf L}_0 \prod_{x=1}^{n-1} {\sf B}_{2x} \times {\sf R}_{2n}
\right) \times
           \prod_{y=1}^n \prod_{x=0}^{n-y} {\sf B}_{y+2x} \,.
 \label{hexT}
\end{eqnarray}

Regarding the handling of the boundary points, a small remark is in
order. In (\ref{hexT}) these have been denoted simply $0$ (on the
left) and $2n+1$ (on the right). In the initial state $|{\rm
  i}\rangle$, both $0$ and $2n+1$ are singletons. After each factor in
the middle product over $y$ the two boundary labels have to be stored,
so that in the final state (\ref{final_state}) the partitions indeed
involve all $6n$ terminals. To avoid introducing a cumbersome
notation, we understand implicitly that this storing is performed when
expanding the product (\ref{hexT}).

\subsubsection{Other lattices} \label{sec:OL}

The $(3,12^2)$ lattice can be handled similarly by rotating ${\sf
  B}_i$ shown in the right part of Figure~\ref{fig:threetwelveTM}
through angle $\pi/2$ clockwise. The left (resp.\ right) boundary
operator ${\sf L}_i$ (resp.\ ${\sf R}_i$) then consists of the four
rightmost (resp.\ five leftmost) edges in the rotated ${\sf B}_i$.
Explicitly we find
\begin{eqnarray}
 {\sf B}_i &=& (v^9 + 6 v^8 + 9 v^7) (iji'j') +
   (v^7 + 3 v^6) \left[ (iji')(j') + (iji')(j) \right] \nonumber \\
 &+& (v^8 + 7 v^7 + (12+q) v^6 + 3 q v^5) \left[ (i)(ji'j') + (i')(ijj') \right]
\nonumber \\
 &+& (v^8 + 8 v^7 + (15+2q) v^6 + 7 q v^5 + q^2 v^4) (ii')(jj') \nonumber \\
 &+& (3 v^6 + (11+q) v^5 + 6 q v^4 + q^2 v^3) (ii')(j)(j') \\
 &+& (3 v^7 + (27+q) v^6 + (56+21q) v^5 + (63q+7q^2) v^4 \nonumber \\
 &  & + (33q^2+q^3) v^3 + 9 q^3 v^2 + q^4 v) (i)(i')(jj') \nonumber \\
 &+& (v^6 + 4 v^5 + q v^4)
     \left[ (i)(j)(i'j') + (ij)(i')(j') + (i)(i'j)(j') + (i')(ij')(j) \right]
\nonumber \\
 &+& (8 v^5 + (40+5q) v^4 + (48q+q^2) v^3 + 27 q^2 v^2 + 8 q^3 v + q^4)
(i)(j)(i')(j') \nonumber
\end{eqnarray}
along with
\begin{eqnarray}
 {\sf L}_i &=& (v^4 + 3 v^3) (ijj') + (v^3 + 4 v^2 + q v) (i)(jj') \nonumber \\
 &+& v^2 \left[ (ij)(j') + (ij')(j) \right] + (3 v + q) (i)(j)(j')
\end{eqnarray}
and
\begin{eqnarray}
 {\sf R}_i &=& (v^5 + 3 v^4) (ii'j) + (v^4 + 4 v^3 + q v^2) \left[ (ij)(i') +
(i)(i'j) \right]
               \nonumber \\
 &+& v^3 (ii')(j) + (v^3 + 8 v^2 + 5 q v + q^2) (i)(j)(i') \,.
\end{eqnarray}
With these expressions for ${\sf B}_i$, ${\sf L}_i$ and ${\sf R}_i$, the
transfer matrix is still given by (\ref{hexT}).

\subsection{Distinguishing 2D, 1D and 0D partitions}
\label{sec:2D1D0D}

We now explain how each partition entering the final state
(\ref{final_state}) can be assigned the correct homotopy (0D, 1D or
2D) in order to make possible the application of the main result
(\ref{eq:2D0D}). The definition of homotopy that we have given in
section~\ref{sec:altdef} is not very practical, because it refers to
the connectivity properties between two arbitrarily separated copies
of the basis, $B_1$ and $B_2$. The purpose of this section is to
provide an operational determination of the homotopy using just intrinsic
properties of $B$.

Each partition of the set of $N$ terminals can be represented as a
planar hypergraph on $N$ vertices, where each part of size $k>1$ in
the partition corresponds to a hyperedge of degree $d = k-1$ in the
hypergraph. Because of the planarity we can obtain yet another
representation as an ordinary graph on $2N$ vertices with precisely
$N$ ordinary ($d=1$) edges. We now detail this construction, which is
completely analogous to a well-known \cite{BaxterKellandWu1976}
equivalence for the partition function of the Potts model defined on a
planar graph $G$ that can be represented, on the one hand, in terms of
Fortuin-Kasteleyn clusters \cite{FK1972} on $G$ and, on the other
hand, as a loop model on the medial graph ${\cal M}(G)$.

The hypergraph can be drawn inside the frame (the outer boundary of
the shaded areas in Figures~\ref{fig:kagomeTM} and
\ref{fig:kagome_hex}) on which the $N$ terminals live. Here we give a
few examples:
\begin{center}
\begin{tikzpicture}[scale=0.5]
 \draw[blue,dashed,line width=0.2ex] (0,0)--(6,0)--(6,6)--(0,6)--cycle;
 \foreach \xpos in {1,3,5}
 \foreach \ypos in {0,6}
   \draw[black,fill] (\xpos,\ypos) circle(1ex);
 \foreach \xpos in {0,6}
 \foreach \ypos in {1,3,5}
   \draw[black,fill] (\xpos,\ypos) circle(1ex);
 \draw[black,line width=0.5ex] (0,5)--(6,3);
 \draw[black,line width=0.5ex] (1,0)--(5,6);
 \draw[black,line width=0.5ex] (0,1) arc(-90:90:1cm);
 \draw[black,line width=0.5ex] (3,0) arc(180:0:1cm);

 \begin{scope}[xshift=10cm]
 \draw[blue,dashed,line width=0.2ex] (0,0)--(6,0)--(6,6)--(0,6)--cycle;
 \foreach \xpos in {1,3,5}
 \foreach \ypos in {0,6}
   \draw[black,fill] (\xpos,\ypos) circle(1ex);
 \foreach \xpos in {0,6}
 \foreach \ypos in {1,3,5}
   \draw[black,fill] (\xpos,\ypos) circle(1ex);
 \draw[black,line width=0.5ex] (0,5)--(6,3);
 \draw[black,line width=0.5ex] (1,0)--(5,6);
 \draw[black,line width=0.5ex] (0,3)--(3.5,3.75);
 \draw[black,line width=0.5ex] (3,0) arc(180:0:1cm);
 \end{scope}

 \begin{scope}[xshift=20cm]
 \draw[blue,dashed,line width=0.2ex] (0,0)--(6,0)--(6,6)--(0,6)--cycle;
 \foreach \xpos in {1,3,5}
 \foreach \ypos in {0,6}
   \draw[black,fill] (\xpos,\ypos) circle(1ex);
 \foreach \xpos in {0,6}
 \foreach \ypos in {1,3,5}
   \draw[black,fill] (\xpos,\ypos) circle(1ex);
 \draw[black,line width=0.5ex] (0,5)--(6,3);
 \draw[black,line width=0.5ex] (1,0)--(5,6);
 \draw[black,line width=0.5ex] (0,3)--(3.5,3.75);
 \draw[black,line width=0.5ex] (5,0)--(3.5,3.75);
 \draw[black,line width=0.5ex] (1,6) arc(180:360:1cm);
 \end{scope}
\end{tikzpicture}
\end{center}
Now place a pair of points slightly shifted on either side of each of
the $N$ terminals.  Draw $N$ edges between these $2N$ points by
``turning around'' the hyperedges and isolated vertices of the
hypergraph. We shall refer to this as the surrounding graph. For each
of the above examples this produces:
\begin{center}
\begin{tikzpicture}[scale=0.5]
 \draw[blue,dashed,line width=0.2ex] (0,0)--(6,0)--(6,6)--(0,6)--cycle;
 \foreach \xpos in {0.7,1.3,2.7,3.3,4.7,5.3}
 \foreach \ypos in {0,6}
   \draw[red,fill] (\xpos,\ypos) circle(0.5ex);
 \foreach \xpos in {0,6}
 \foreach \ypos in {0.7,1.3,2.7,3.3,4.7,5.3}
   \draw[red,fill] (\xpos,\ypos) circle(0.5ex);
 \draw[black!30,line width=0.5ex] (0,5)--(6,3);
 \draw[black!30,line width=0.5ex] (1,0)--(5,6);
 \draw[black!30,line width=0.5ex] (0,1) arc(-90:90:1cm);
 \draw[black!30,line width=0.5ex] (3,0) arc(180:0:1cm);
 \draw[red,line width=0.3ex] (0,0.7) arc(-90:90:1.3cm);
 \draw[red,line width=0.3ex] (0,1.3) arc(-90:90:0.7cm);
 \draw[red,line width=0.3ex] (0.7,0)--(2.75,3.15) arc(-30:80:5mm)--(0,4.7);
 \draw[red,line width=0.3ex] (0,5.3)--(3,4.3) arc(260:330:7mm)--(4.7,6);
 \draw[red,line width=0.3ex] (0.7,6) arc(180:360:0.3cm);
 \draw[red,line width=0.3ex] (2.7,6) arc(180:360:0.3cm);
 \draw[red,line width=0.3ex] (5.3,6)--(4.2,4.3) arc(150:260:3mm)--(6,3.3);
 \draw[red,line width=0.3ex] (6,5.3) arc(90:270:0.3cm);
 \draw[red,line width=0.3ex] (6,2.7)--(4,3.4) arc(70:130:5mm)--(1.3,0);
 \draw[red,line width=0.3ex] (6,1.3) arc(90:270:0.3cm);
 \draw[red,line width=0.3ex] (5.3,0) arc(0:180:1.3cm);
 \draw[red,line width=0.3ex] (4.7,0) arc(0:180:0.7cm);

 \begin{scope}[xshift=10cm]
 \draw[blue,dashed,line width=0.2ex] (0,0)--(6,0)--(6,6)--(0,6)--cycle;
 \foreach \xpos in {0.7,1.3,2.7,3.3,4.7,5.3}
 \foreach \ypos in {0,6}
   \draw[red,fill] (\xpos,\ypos) circle(0.5ex);
 \foreach \xpos in {0,6}
 \foreach \ypos in {0.7,1.3,2.7,3.3,4.7,5.3}
   \draw[red,fill] (\xpos,\ypos) circle(0.5ex);
 \draw[black!30,line width=0.5ex] (0,5)--(6,3);
 \draw[black!30,line width=0.5ex] (1,0)--(5,6);
 \draw[black!30,line width=0.5ex] (0,3)--(3.5,3.75);
 \draw[black!30,line width=0.5ex] (3,0) arc(180:0:1cm);
 \draw[red,line width=0.3ex] (0,0.7) arc(-90:90:0.3cm);
 \draw[red,line width=0.3ex] (0.7,0)--(2.5,2.7) arc(-30:120:3.5mm)--(0,2.7);
 \draw[red,line width=0.3ex] (0,3.3)--(2,3.7) arc(-80:80:2mm)--(0,4.7);
 \draw[red,line width=0.3ex] (0,5.3)--(3,4.3) arc(260:330:7mm)--(4.7,6);
 \draw[red,line width=0.3ex] (0.7,6) arc(180:360:0.3cm);
 \draw[red,line width=0.3ex] (2.7,6) arc(180:360:0.3cm);
 \draw[red,line width=0.3ex] (5.3,6)--(4.2,4.3) arc(150:260:3mm)--(6,3.3);
 \draw[red,line width=0.3ex] (6,5.3) arc(90:270:0.3cm);
 \draw[red,line width=0.3ex] (6,2.7)--(4,3.4) arc(70:130:5mm)--(1.3,0);
 \draw[red,line width=0.3ex] (6,1.3) arc(90:270:0.3cm);
 \draw[red,line width=0.3ex] (5.3,0) arc(0:180:1.3cm);
 \draw[red,line width=0.3ex] (4.7,0) arc(0:180:0.7cm);
 \end{scope}

 \begin{scope}[xshift=20cm]
 \draw[blue,dashed,line width=0.2ex] (0,0)--(6,0)--(6,6)--(0,6)--cycle;
 \foreach \xpos in {0.7,1.3,2.7,3.3,4.7,5.3}
 \foreach \ypos in {0,6}
   \draw[red,fill] (\xpos,\ypos) circle(0.5ex);
 \foreach \xpos in {0,6}
 \foreach \ypos in {0.7,1.3,2.7,3.3,4.7,5.3}
   \draw[red,fill] (\xpos,\ypos) circle(0.5ex);
 \draw[black!30,line width=0.5ex] (0,5)--(6,3);
 \draw[black!30,line width=0.5ex] (1,0)--(5,6);
 \draw[black!30,line width=0.5ex] (0,3)--(3.5,3.75);
 \draw[black!30,line width=0.5ex] (5,0)--(3.5,3.75);
 \draw[black!30,line width=0.5ex] (1,6) arc(200:340:10.7mm);
 \draw[red,line width=0.3ex] (0,0.7) arc(-90:90:0.3cm);
 \draw[red,line width=0.3ex] (0.7,0)--(2.5,2.7) arc(-30:120:3.5mm)--(0,2.7);
 \draw[red,line width=0.3ex] (0,3.3)--(2,3.7) arc(-80:80:2mm)--(0,4.7);
 \draw[red,line width=0.3ex] (0,5.3)--(3,4.3) arc(260:330:7mm)--(4.7,6);
 \draw[red,line width=0.3ex] (0.7,6) arc(200:340:14mm);
 \draw[red,line width=0.3ex] (1.3,6) arc(200:340:7.5mm);
 \draw[red,line width=0.3ex] (5.3,6)--(4.2,4.3) arc(150:260:3mm)--(6,3.3);
 \draw[red,line width=0.3ex] (6,5.3) arc(90:270:0.3cm);
 \draw[red,line width=0.3ex] (6,2.7)--(4.3,3.3) arc(70:210:2mm)--(5.3,0);
 \draw[red,line width=0.3ex] (6,1.3) arc(90:270:0.3cm);
 \draw[red,line width=0.3ex] (4.7,0)--(3.7,2.6) arc(10:150:3mm)--(1.3,0);
 \draw[red,line width=0.3ex] (3.3,0) arc(0:180:0.3cm);
 \end{scope}
\end{tikzpicture}
\end{center}

The embedding of $B$ is defined by identifying points on opposing
sides of the frame (to produce the twisted embeddings we further shift
the points on one of the sides cyclically before imposing the
identification).  Let $\ell$ be the number of loops in the surrounding
graph. The partition is of the 1D type if and only if one or more of
these loops is non-homotopic to a point. To determine whether this is
the case it suffices to ``follow'' each loop until one comes back to
the starting point, and determine whether the total signed
displacement in the $x$ and $y$ directions is non-zero.%
\footnote{For the straight embedding one can more simply determine
  whether the signed winding number with respect to any of the two
  periodic directions is non-zero.}
Using this method one sees that the middle partition in the above
three examples is of the 1D type.

If all loops on the surrounding graph have trivial homotopy, one
can use the Euler relation to determine whether the partition is
of the 0D or 2D type. Namely, let $E$ be the sum of all degrees
of the hyperedges in the hypergraph; let $C$ (resp.\ $V$) be the
number of connected components (resp.\ vertices) in the hypergraph
after the identification of opposing sides. Then the combination
\begin{equation}
 \chi = E + 2 C - V - \ell
\end{equation}
equals 0 (resp.\ 2) if the partition is of the 0D (resp.\ 2D) type.

For instance, for the leftmost example we have $E = 3 + 1 + 1 = 5$, $C
= 1$, $V = 6$, and $\ell = 1$, whence $\chi = 0$. And for the
rightmost example one finds $E = 5 + 1 = 6$, $C = 2$, $V = 6$, and
$\ell = 2$, whence $\chi = 2$.

\subsection{Number of connected components containing a terminal}
\label{sec:q_power}

To finish the computation of the factor $q^{k(A)-1}$ in (\ref{eq:vq_factors}),
it remains to count the number of connected components (minus one) containing at
least one terminal.
The terminals are nothing but the points in the partition describing the final
state (\ref{final_state}).
But before counting we need to identify the matching terminals, as shown in
Figures~\ref{fig:squarekagome}--\ref{fig:hexkagome}.

The algebraic formulation of this counting procedure is simple. Let the number
of terminals
be $2N$, with $N=2n$ (resp.\ $N=3n$) for the square (resp.\ hexagonal) bases. 
Matching terminals are labelled $\{i_k,j_k\}$, for $k=1,2,\ldots,N$.
The operator ${\sf J}$ identifying all matching pairs of terminals is then
expressed in terms
of join operators ${\sf J}_{ij}$ as
\be
 {\sf J} = \prod_{k=1}^N {\sf J}_{i_k,j_k} \,.
\ee
To multiply the weights by the correct power of $q$ we finally apply
\be
 {\sf D} = q^{-1} \prod_{k=1}^N {\sf D}_{i_k} {\sf D}_{j_k} \,.
\ee

More precisely, let $|{\rm f}\rangle_{2D}$ denote the terms in the final state
(\ref{final_state}) whose partitions
are of 2D topology. Then
\be
 {\sf D} {\sf J} |{\rm f}\rangle_{2D} = W(2D;B) \,
(i_1)(i_2)\cdots(i_N)(j_1)(j_2)\cdots(j_N) \,,
\ee
i.e., one of the weights needed in (\ref{eq:2D0D}) times the all-singleton
state. We similarly obtain $W(0D;B)$
from $|{\rm f}\rangle_{0D}$.

Let us illustrate this procedure by a small example. The final state $|{\rm
f}\rangle$ corresponding to
the $1 \times 1$ square basis for the kagome lattice can be read directly off
the right-hand side of
(\ref{newBkagome}). The unique partition having 2D topology is $(iji'j')$. The
partitions having 0D
topology are $(ii')(j)(j')$, $(i)(i')(jj')$, $(i)(j)(i')(j')$, $(i)(j)(i'j')$
and $(ij)(i')(j')$. The corresponding terms
in the sum (\ref{newBkagome}) then define $|{\rm f}\rangle_{2D}$ and $|{\rm
f}\rangle_{0D}$.
Applying the operator ${\sf D} {\sf J}$ as above, we infer that
\ba
 W(2D;B) &=& v^6 + 6 v^5 + 9 v^4 \,, \\
 W(0D;B) &=& 2 \times (v^3 + 5 v^2 + q v) + (4 v + q) \times q + 2 \times v^2
\,.
\ea
According to (\ref{eq:2D0D}) the critical polynomial is then
\be
 P_B(q,v) = v^6 + 6 v^5 + 9 v^4 - 2 q v^3 - 12 q v^2 - 6 q^2 v - q^3 \,.
 \label{kagome_wu}
\ee
The expression $P_B(q,v) = 0$ is precisely the approximation to the
kagome-lattice critical manifold
found by Wu \cite{Wu79}.

\section{Results}
\label{sec:results}

Using the transfer matrix approach we have computed the critical polynomials
$P_B(q,v)$ for the
kagome, $(4,8^2)$ and $(3,12^2)$ lattices; see Figure~\ref{fig:bondlattices}.
With the $n \times n$ square bases, the computations
were possible for $n \le 4$, both with straight and twisted embeddings.
This is a considerable improvement over the contraction-deletion method,
where the $2 \times 3$ rectangular basis (with 36 edges for the kagome lattice)
was the furthest we could go \cite{Jacobsen12}. With the hexagonal bases, the
transfer matrix
computations were possible for $n \le 2$. For selected integer values of $q$,
the case $n=3$ was within
reach as well, albeit with large parallel computations; the $n=3$ results for percolation ($q=1$) have been reported in
\cite{SJ12}.

The critical polynomials that we have obtained are very large. To be precise,
let $|V|$ (resp.\ $|E|$)
denote the number of vertices (resp.\ edges) in the basis, with the convention
that a pair of matching
terminals counts as a single vertex. Then $P_B(q,v)$ is a polynomial of degree
$|V|$ in the $q$
variable, and of degree $|E|$ in the $v$ variable,
with very large integer coefficients (more than 100 digits in some cases). Obviously it is out
of the question to make
these polynomials appear in print. However, all the polynomials are
collected in the text
file {\tt JS12.m} which is available in electronic form as supplementary
material to this paper.%
\footnote{This file can be processed by {\sc Mathematica} or---maybe
  after minor changes of formatting---by any symbolic computer algebra
  program of the reader's liking.}
The printed version contains only selected values of the roots $v_c$,
rounded to 15 digit numerical precision, and plots of the curves $P_B(q,v) = 0$
in the real $(q,v)$ plane (see section~\ref{sec:phase_diag}).

In this section we analyse the approximations to the critical temperatures,
$v_c$,
obtained by solving $P_B(q,v) = 0$ in the ferromagnetic regime ($v>0$) for
selected
integer values of $q$ of practical interest, namely $q=2,3,4$. The corresponding
results
for $q=1$ have already appeared in \cite{SJ12}. Then, in
section~\ref{sec:phase_diag},
we extend the discussion to the full phase diagram in the real $(q,v)$ plane.

\subsection{Kagome lattice}

For the kagome lattice, we considered two families of bases: square (see
section~\ref{sec:sq_bases}) and hexagonal (see section~\ref{sec:hex_bases}).

\subsubsection{Square bases}

The $n \times n$ square bases with straight and twisted embeddings are
shown in Figure~\ref{fig:squarekagome}. They contain $|V| = 3 n^2$ vertices
and $|E| = 6 n^2$ edges.  We have obtained the critical polynomials for $n
\le 4$ and twist $k \le \lfloor n/2 \rfloor$.

For $q=2$ all those polynomials factorise, shedding the small factor
\be
 v^4 + 4 v^2 - 8v - 8 \,.
 \label{eq:ising_factor_kagome}
\ee
This factorisation is expected \cite{Jacobsen12}, since the Ising model is
exactly solvable. The unique positive root of (\ref{eq:ising_factor_kagome})
reads
\be
 v_c = \sqrt{3 + 2 \sqrt{3}} - 1 = 1.542\,459\,756\cdots \,,
\ee
in agreement with the exact solution \cite{KanoNaya53}.

\begin{table}
\begin{center}
\begin{tabular}{c|c|l|l}
 $n$ & twist & $v_c$ for $q=3$ & $v_c$ for $q=4$ \\
\hline \hline
1 & 0 & 1.876\,269\,208\,345\,761 & 2.155\,842\,236\,513\,638 \\
\hline
2 & 0 & 1.876\,439\,754\,302\,881 & 2.156\,207\,452\,990\,795 \\
\  & 1 & 1.876\,439\,754\,302\,881 & 2.156\,207\,452\,990\,795 \\
\hline
3 & 0 & 1.876\,456\,916\,196\,415 & 2.156\,247\,598\,338\,124 \\
\  & 1 & 1.876\,456\,775\,276\,423 & 2.156\,247\,104\,437\,170 \\
\hline
4 & 0 & 1.876\,458\,994\,003\,462 & 2.156\,252\,880\,154\,217 \\
\  & 1 & 1.876\,458\,930\,268\,948 & 2.156\,252\,686\,506\,350 \\
\  & 2 & 1.876\,458\,896\,874\,446 & 2.156\,252\,553\,917\,008 \\
\hline \hline
Final & & 1.876\,459\,0 \,(5) & 2.156\,253 \,(1) \\
\hline
Numerics & Ref.~\cite{Jacobsen12} & 1.876\,459 \,(2) & 2.156\,252\,(2) \\
                  & Ref.~\cite{Ding10}          & 1.876\,458 \,(3) &
2.156\,20\,(5) \\
\end{tabular}
\caption{Predictions for the critical temperature $v_c$ of the Potts model on
the kagome lattice,
 obtained from the $n \times n$ square bases with various twists. Extrapolation
to $n \to \infty$
 leads to the final results quoted. For comparison we show two sets of recent
numerical results \cite{Jacobsen12,Ding10}.}
\label{tab:squarekagome}
\end{center}
\end{table}

The positive roots of $P_B(q,v)$ for $q=3$ and $q=4$ are given in
Table~\ref{tab:squarekagome}. Note that the results for $(n,k) =
(2,0)$ and $(2,1)$ are identical for this lattice; but otherwise the
critical polynomial does depend on $k$.
Based on the results for finite $n$ we suggest the final values
shown in the bottom of Table~\ref{tab:squarekagome}.
They are compatible with the results for the largest ($n=4$) basis for any $k$,
and the indicative error bar has been obtained by comparing the
variation of the results for various twists, and by a crude analysis%
\footnote{Unfortunately the number of data points is too small to allow the
application
of really efficient extrapolation methods, such as the Bulirsch-Stoer
algorithm.}
of the the finite-size effects in $n$ with zero twist.
Our final values agree with---and are marginally more precise than---two
sets of recent numerical results, obtained respectively from the
crossings of effective critical exponents \cite{Ding10} and from the
maximum of the effective central charge \cite{Jacobsen12}.

\subsubsection{Hexagonal bases}

The hexagonal bases of size $n$ are shown in
Figure~\ref{fig:hexkagome}. They contain $|V| = 9 n^2$ vertices and $|E| = 18
n^2$
edges. As discussed in section~\ref{sec:hex_bases}, these bases better
respect the rotational symmetry of the lattice, and hence we expect
the results to be more precise than those with the square bases for a
given number of edges. We have obtained the full critical polynomials $P_B(q,v)$
for $n
\le 2$. For $n=3$ we have also obtained polynomials%
\footnote{For $n=3$, the basis has $18$ terminals and a very
large calculation is necessary. This was done in parallel on Lawrence Livermore
National Laboratory's Cab supercomputer, utilizing $2046$ processors, each $2.6$
GHz, for about $30$ hours. In brief, the parallelism is achieved by distributing the state vector over the processors. The challenge is then ensuring that data is communicated properly upon application of the ${\sf B}$,
${\sf L}$ and ${\sf R}$ operators.} $P_B(q_0,v)$ in the $v$
variable
for the integer values $q_0=3$ and 4  (see also \cite{SJ12} for the case $q_0 =
1$).

For $q=2$ the polynomials again yield the exact result, since they
invariably contain the factor (\ref{eq:ising_factor_kagome}). Results
for $q=3$ and $q=4$ are given in Table~\ref{tab:hexkagome}.

\begin{table}
\begin{center}
\begin{tabular}{c|l|l}
 $n$ & $v_c$ for $q=3$ & $v_c$ for $q=4$ \\
\hline \hline
1 & 1.876\,456\,753\,812\,346 & 2.156\,240\,076\,964\,790 \\
\hline
2 & 1.876\,458\,831\,984\,027 & 2.156\,252\,155\,471\,646 \\
\hline
3 & 1.876\,459\,465\,279\,159 & 2.156\,254\,172\,414\,827 \\
\hline \hline
Final & 1.876\,459\,7 \,(2) & 2.156\,254\,5 \,(3) \\
\end{tabular}
\caption{Predictions for $v_c$ on the kagome lattice, using the hexagonal bases
of size $n$.}
\label{tab:hexkagome}
\end{center}
\end{table}

It should be noticed that the $n=2$ results with a hexagonal basis (72 edges)
have a precision comparable to that of the $n=4$ results with a square basis (96
edges).
This presumably indicates that the hexagonal-basis results converge at a faster
rate
than the square-basis results, since they respect better the rotational symmetry
of the kagome lattice.

\subsection{$(4,8^2)$ lattice}

\begin{figure}
\begin{center}
\includegraphics{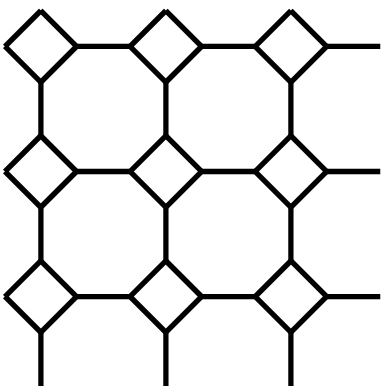}
\caption{The $3 \times 3$ square basis, with unspecified embedding, for the
$(4,8^2)$ lattice.}
\label{fig:FE3x3}
\end{center}
\end{figure}

We computed the critical polynomials for the $n \times n$ square bases
on the $(4,8^2)$ lattice (see Figure~\ref{fig:FE3x3}). They contain $|V| = 4
n^2$ vertices
and $|E| = 6 n^2$ edges. As this graph
does not have the kagome lattice's hexagonal symmetry, there are no
corresponding hexagonal bases.

Results for $n \le 4$ are given in
Table~\ref{tab:foureight}, with the twists $k \le \lfloor n/2 \rfloor$
defined identically to the kagome case.  Note that the cases $(n,k) =
(2,0)$ and $(2,1)$ now produce different results. Unfortunately we are
not aware of any numerical results with which to compare our results.

\begin{table}
\begin{center}
\begin{tabular}{c|c|l|l}
 $n$ & twist & $p_c$ \\
\hline \hline
1 & 0 & 3.742\,119\,707\,930\,615 & 4.367\,630\,831\,288\,119 \\
\hline
2 & 0 & 3.742\,406\,812\,389\,425 & 4.368\,211\,338\,019\,044 \\
\  & 1 & 3.742\,464\,337\,713\,004 & 4.368\,322\,400\,865\,888 \\
\hline
3 & 0 & 3.742\,474\,558\,548\,594 & 4.368\,344\,356\,164\,380 \\
\  & 1 & 3.742\,485\,404\,327\,335 & 4.368\,364\,878\,291\,854 \\
\hline
4 & 0 & 3.742\,488\,803\,421\,387 & 4.368\,371\,674\,728\,465 \\
\  & 1 & 3.742\,491\,065\,678\,059 & 4.368\,375\,885\,780\,634 \\
\  & 2 & 3.742\,492\,580\,864\,574 & 4.368\,378\,693\,163\,071 \\
\hline \hline
Final & & 3.742\,489 \, (4) & 4.368\,372 (7) \\
\end{tabular}
\caption{Predictions for $v_c$ on the $(4,8^2)$ lattice, using the $n \times n$
square bases with various twists.
 Extrapolation to $n \to \infty$ leads to the final results quoted.}
\label{tab:foureight}
\end{center}
\end{table}

For $q=2$ all the polynomials factorise, shedding the small factor
\be
  v^4 - 6 v^2 - 8 v - 4 \,,
\ee
whose unique positive root
\be
 v_c = \frac{1+ \sqrt{5 + 4 \sqrt{2}}}{\sqrt{2}} = 3.015\,445\,388\cdots
\ee
provides the exact critical point of the Ising model on the $(4,8^2)$ lattice \cite{Hurst63}.

\subsection{$(3,12^2)$ lattice}

The $(3,12^2)$ lattice bears more than a passing resemblance to the
kagome lattice. Employing the analogous $n \times n$ square bases and
twists, we find the results in Table~\ref{tab:squareTT}. 
Note that these bases contain $|V| = 6 n^2$ vertices
and $|E| = 9 n^2$ edges.

\begin{table}
\begin{center}
\begin{tabular}{c|c|l|l}
 $n$ & twist & $v_c$ for $q=3$ & $v_c$ for $q=4$ \\
\hline \hline
1 & 0 & 5.033\,022\,514\,872\,745 & 5.857\,394\,827\,983\,648 \\
\hline
2 & 0 & 5.033\,072\,313\,070\,887 & 5.857\,498\,027\,767\,977 \\
\  & 1 & 5.033\,072\,313\,070\,887 & 5.857\,498\,027\,767\,977 \\
\hline
3 & 0 & 5.033\,077\,636\,920\,826 & 5.857\,509\,929\,206\,085 \\
\  & 1 & 5.033\,077\,582\,117\,669 & 5.857\,509\,766\,966\,587 \\
\hline
4 & 0 & 5.033\,078\,299\,711\,932 & 5.857\,511\,525\,138\,037 \\
\  & 1 & 5.033\,078\,277\,586\,327 & 5.857\,511\,464\,406\,024 \\
\  & 2 & 5.033\,078\,264\,247\,232 & 5.857\,511\,420\,811\,147 \\
\hline \hline
Final & & 5.033\,078\,3 \,(2) &  5.857\,511\,5 \,(5) \\
\hline
Numerics & Ref.~\cite{Ding10}  & 5.033\,077 \,(3) & 5.857\,497 \,(3) \\
\end{tabular}
\caption{Predictions for $v_c$ on the $(3,12^2)$ lattice, using the $n \times n$
square bases with various twists.
 Extrapolation to $n \to \infty$ leads to the final results quoted. For
comparison we show also some recent numerical results \cite{Ding10}.}
\label{tab:squareTT}
\end{center}
\end{table}

Our final results can be compared with the recent numerical results of
\cite{Ding10}, obtained
from the crossings of effective critical exponents. It transpires that the error
bar on the result
for $v_c(q=4)$ reported in \cite{Ding10} is underestimated by at least a factor
of five, i.e., it
should have read something like $5.857\,497 \,(15)$. It follows that both for
$q=3$ and $q=4$,
the accuracy of our final result improves on that of \cite{Ding10} by more than
an order of magnitude.

It should be stressed that loss of precision and problems estimating proper
error bars are very
common in studies of the $q=4$ state Potts model. This is due to the presence of
a marginally
irrelevant scaling operator in the conformal field theory that describes the
continuum limit, which has the effect
of introducing logarithmic corrections to scaling. It is remarkable that the
method of critical polynomials
appears to be insensitive to such logarithmic corrections, yielding results that
are of comparable
precision for $q=3$ and $q=4$.

Just as in previous cases, we find for $q=2$ that all the polynomials factorise,
shedding now the small factor
\be
  v^4 - 2 v^3 - 6 v^2 - 8 v - 8 \,.
\ee
Its unique positive root
\be
 v_c = \frac12 \left( 1 + \sqrt{3} + \sqrt{2(6 + 5 \sqrt{3})} \right) =
4.073\,446\,135\cdots
 \ee
provides the exact critical point of the Ising model on the $(3,12^2)$ lattice \cite{Syozi55}.

\begin{table}
\begin{center}
\begin{tabular}{c|l|l}
 $n$ & $v_c$ for $q=3$ & $v_c$ for $q=4$ \\
\hline \hline
1 & 5.033\,076\,898\,972\,026 & 5.857\,506\,572\,441\,733 \\
\hline
2 & 5.033\,078\,231\,476\,569 & 5.857\,511\,281\,996\,374 \\
\hline
3 & 5.033\,078\,451\,436\,561 & 5.857\,511\,917\,242\,462 \\
\hline \hline
Final & 5.033\,078\,49 \,(4) &  5.857\,512\,00 \,(8) \\
\end{tabular}
\caption{Predictions for $v_c$ on the $(3,12^2)$ lattice, using the hexagonal
bases of size $n$.}
\label{tab:hexTT}
\end{center}
\end{table}

Results with the hexagonal bases are shown in
Table~\ref{tab:hexTT}.%
\footnote{The $n=3$ calculations required about 30 hours on 4092
processors, each 2.6 GHz.}
These bases contain $|V| = 18 n^2$ vertices and $|E| = 27 n^2$ edges.
Just as for the kagome lattice, we see that the $n=2$ results with a hexagonal
basis (108 edges)
have a precision comparable to that of the $n=4$ results with a square basis
(144 edges).

\section{Phase diagrams}
\label{sec:phase_diag}

Obviously the critical polynomials $P_B(q,v)$ contain much more information than
the sporadic critical
points reported in section~\ref{sec:results}. We shall now see how the roots of
$P_B(q,v)$ in the real $(q,v)$
plane yield detailed information about the whole critical manifold of the Potts
model on the relevant lattice.
The behaviour in the antiferromagnetic region $v < 0$ turns out to be
particularly rich.
We limit the investigation to the half-plane $q \ge 0$. Results for the kagome
lattice, using bases smaller
than those reported here, have already appeared in \cite{Jacobsen12}.

\subsection{Square lattice}

The square lattice is the only lattice for which the critical manifold of the
Potts model is analytically understood 
in the entire real $(q,v)$ plane. It thus constitutes an interesting benchmark
case, where we can confront
the solutions of $P_B(q,v) = 0$ with known results about the phase diagram.
Therefore we review some
pertinent facts about the square-lattice Potts model before discussing the three
lattices---viz., the $(4,8^2)$, kagome
and $(3,12^2)$ lattices---for which no exact solutions are available.

The square-lattice Potts model is exactly solvable on the self-dual curve $v^2 -
q = 0$ \cite{Baxter73} as well as
on the antiferromagnetic manifold $v^2 + 4 v + q = 0$ \cite{Baxter82}. The exact
solvability manifests itself in
the fact \cite{Jacobsen12} that, for any choice of the basis, $P_B(q,v)$
contains the factors $(v^2-q)(v^2+4v+q)$.
These curves are also known to be loci of phase transitions, i.e., they are
contained in the critical manifold.
The phase transitions are second order for $0 \le q \le 4$ and first order for
$q > 4$ \cite{Baxter73}.
In the second-order regime, the critical exponent corresponding to a perturbation
in the temperature variable $v$
is known by a variety of techniques \cite{Baxter73,Baxter82,Saleur91,JacSal06}.
In particular, the temperature
perturbation is irrelevant (in the renormalisation group sense) along the
critical curve $v = -\sqrt{q}$. It follows
that, for any fixed $0 \le q \le 4$, all the points satisfying $-2-\sqrt{4-q} <
v < -2+\sqrt{4-v}$ will flow to the fixed
point $v = -\sqrt{q}$. In other words, the physics inside the region bounded by
the two mutually dual antiferromagnetic
transition curves is temperature independent. This region is known as the
Berker-Kadanoff phase \cite{Saleur91}.

Another important fact is that the Potts model on any lattice possesses a
quantum group symmetry \cite{PasquierSaleur90}.
This symmetry implies massive cancellations among transfer matrix eigenvalues
(or between representations of
the Virasoro algebra in the continuum limit) whenever $q$ is equal to a
so-called Beraha number
\be
 B_k = \left( 2 \cos (\pi / k) \right)^2 \,, \qquad \mbox{with } k=2,3,4,\ldots
\,.
 \label{eq:Beraha}
\ee
The cancellations for $q=B_k$ make it possible to reformulate the Potts model as
an RSOS height model \cite{ABF84}
with strictly local Boltzmann weights, i.e., to dispose of the non-locality that
is inherent to the factor $q^{k(A)}$ appearing
in the generic (i.e., valid for any real value of $q$) formulation
(\ref{FK_repr}).
In general, such cancellations are relevant only for the ``fine structure''
of the model, but in the Berker-Kadanoff phase they impact the dominant term in
the partition function, making
possible further phase transitions. The independence on $v$ implies that the
effect on the phase diagram
is the formation of a vertical ray in the $(q,v)$ plane, with $q=B_k$.

The issue of dominance depends on the choice
of boundary conditions, which therefore determines which $B_k$ are the loci of
phase transitions.
In \cite{Salas06} it was argued from results of conformal field
theory---and checked numerically---that with cyclic boundary
conditions (free in one lattice direction and periodic in the other)
partition function zeros condense along vertical rays with $k \in 2 \N$. The
corresponding result
for toroidal boundary conditions \cite{Salas07} (periodic in both
lattice directions) is that vertical rays occur only for $k=4$ and
$k=6$. It is not completely clear how the results of \cite{Salas06,Salas07} 
would be reflected by the roots of the critical polynomial $P_B(q,v)$.
Because of the identification of opposite terminals in the
embeddings of the basis $B$ that we have used, it might be that the
case of toroidal boundary conditions \cite{Salas07} is most relevant
in the present context. In any case, we certainly expect the Beraha numbers
(\ref{eq:Beraha}) to play an important role in the Berker-Kadanoff phase
\cite{Saleur91}.

\begin{figure}
\begin{center}
\includegraphics[width=12cm]{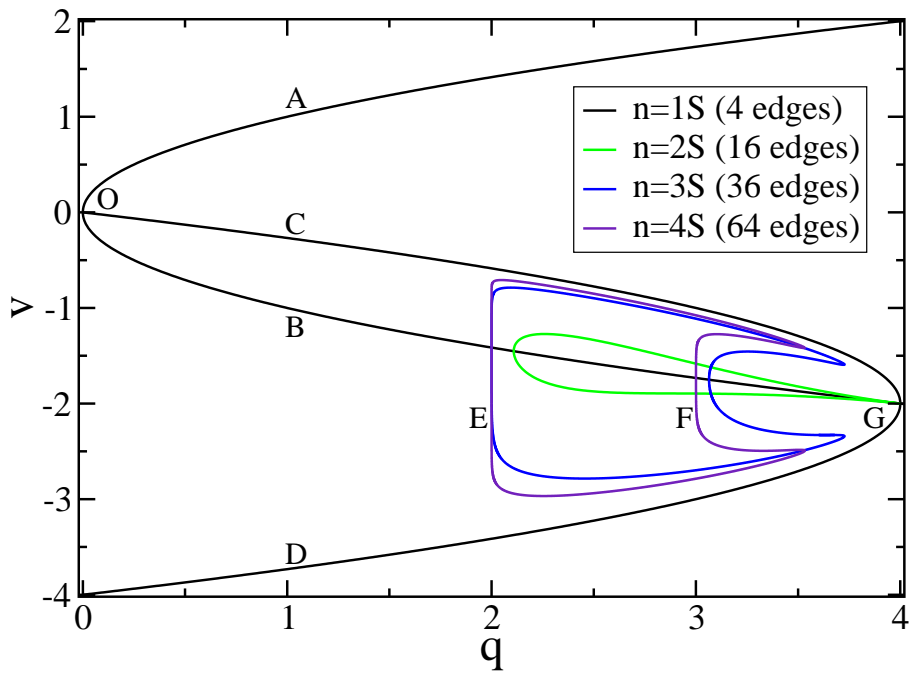}
\caption{Roots of $P_B(q,v)$ for the Potts model on the square lattice, using $n
\times n$ square bases. The
bases are labelled $n$S in the figure legend (S stands for ``square'').
The letters appearing in the figure are explained in the main text.}
\label{fig:sq1}
\end{center}
\end{figure}

To get a better idea about what to expect for other lattices, we show in
Fig.~\ref{fig:sq1} the manifolds $P_B(q,v) = 0$ for the square-lattice
Potts model with $n \times n$ square bases of size $n \le 4$.
The critical polynomials are obtained within the transfer
matrix formalism of section~\ref{sec:tm} by using the building block
\be
 {\sf B}_i = {\sf H}_i {\sf V}_i {\sf V}_{i+1} {\sf H}_i \,.
\ee
The small factors (\ref{sq_latt_cc}), appearing in each of the $P_B(q,v)$,
produce the selfdual critical curves $v^2 - q = 0$ (denoted A and B) and the
dual pair
of antiferromagnetic transition curves $v^2 + 4 v + q = 0$ (denoted C and D).
The remaining large factor in $P_B(q,v)$, of degree $|E|-4$ in the $v$ variable,
produces dual pairs of curves inside the C and D curves, in
the form of ``bubbles'' to the left of the point $(q,v) = (4,-2)$, denoted G. 
Each
bubble intersects the self-dual transition curve $v = -\sqrt{q}$ (denoted B) in
two
points. The corresponding $q$ values, $q_{\rm c}^{(1)}$ and $q_{\rm c}^{(2)}$,
are given in Table~\ref{tab:qc}. It appears that they converge very fast to
$B_4 = 2$ and $B_6 = 3$ upon increasing $n$. Figure~\ref{fig:sq1} provides
convincing
evidence that in the limit $n \to \infty$ each of these two points will be part
of a vertical
ray (E and F) extending between the antiferromagnetic curves (C and D).

While this is in line with the general expectations outlined above, the possible
connexion between $P_B(q,v)$ and the studies \cite{Salas06,Salas07} of partition
function zeros
remains rather indirect. In particular, it remains an open question at this
stage whether
yet larger bases might lead to the formation of vertical rays at other
Beraha numbers than $B_4$ and $B_6$.
\begin{table}[b]
\begin{center}
\begin{tabular}{c|ll}
 $n$ & $q_{\rm c}^{(1)}$ & $q_{\rm c}^{(2)}$ \\
\hline \hline
2 & 2.032\,815\,790\,358\,187 & 4.000\,000\,000\,000\,000 \\
3 & 2.000\,010\,742\,629\,917 & 3.064\,263\,890\,473\,626 \\
4 & 2.000\,000\,000\,040\,606 & 3.000\,370\,123\,813\,456 \\
\hline
$\infty$ & 2 & 3 \\
\end{tabular}
\caption{Crossings of the ``bubbles'' in Figure~\ref{fig:sq1} with the critical
curve $v=-\sqrt{q}$.}
\label{tab:qc}
\end{center}
\end{table}
\subsection{$(4,8^2)$ lattice}

In the following three subsections we discuss the lattices which are the subject
of this study. We have arranged
them in order of increasing complexity of their critical manifolds. We begin
with the $(4,8^2)$ lattice, whose
phase diagram turns out to be very similar to that of the square lattice.

\begin{figure}
\begin{center}
\includegraphics[width=12cm]{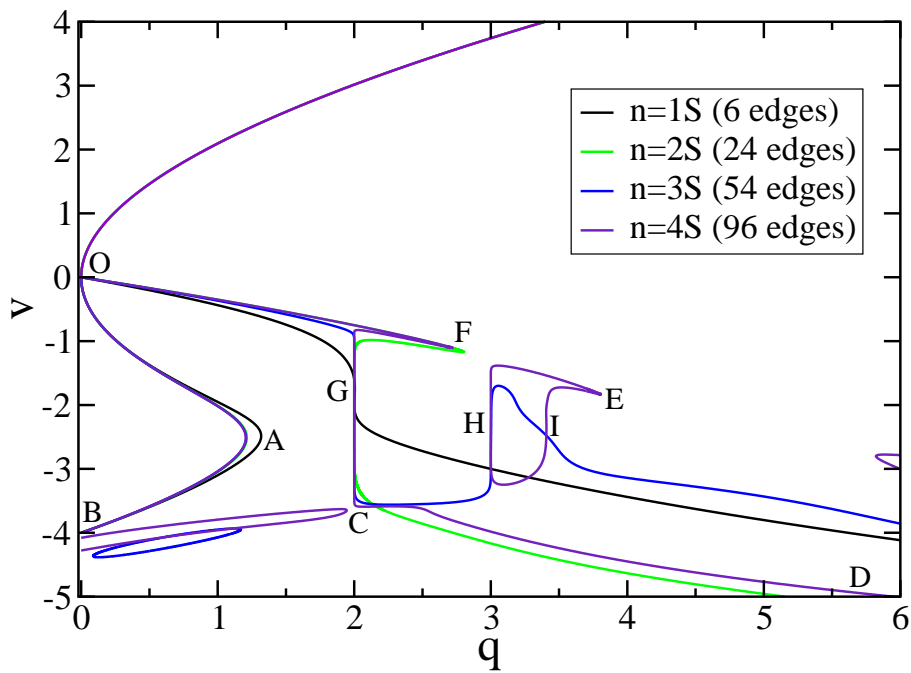}
\caption{Roots of $P_B(q,v)$ for the Potts model on the $(4,8^2)$ lattice, using
$n \times n$ square bases.}
\label{fig:fe1}
\end{center}
\end{figure}

\begin{figure}
\begin{center}
\includegraphics[width=12cm]{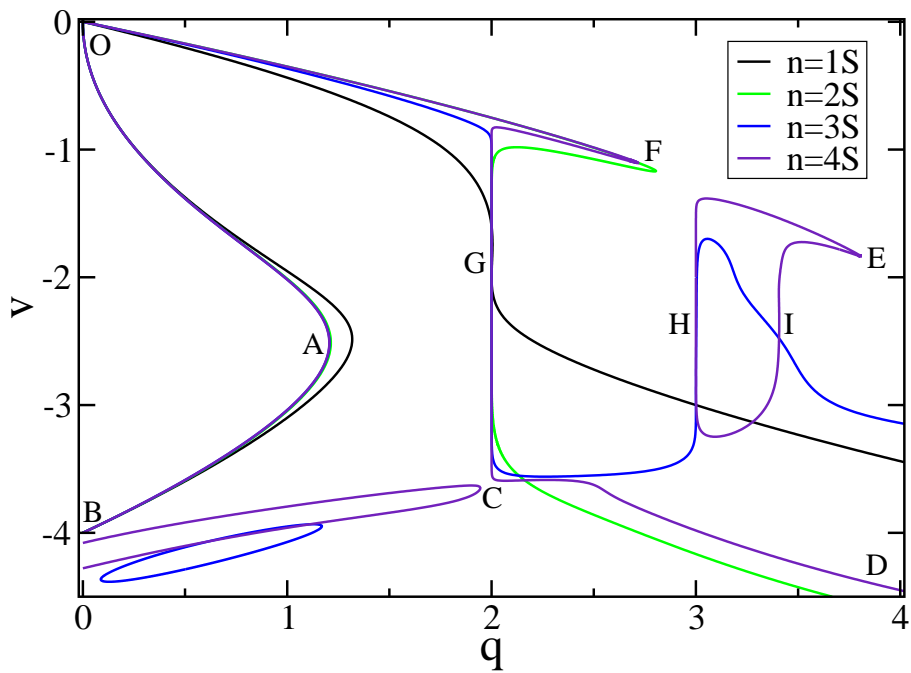}
\caption{Close-up on the antiferromagnetic region of Figure~\ref{fig:fe1}.}
\label{fig:fe2}
\end{center}
\end{figure}

The roots of $P_B(q,v)$ for square bases of size $n \le 4$ are shown in
Figure~\ref{fig:fe1}. In the ferromagnetic
region the curves are indistinguishable on the scale of the figure (see
section~\ref{sec:results} for details).
However, the close-up on the antiferromagnetic region, presented in
Figure~\ref{fig:fe2}, reveals
considerable finite-size effects. The question naturally arises which parts of
these curves provide reliable
information about the critical manifold in the thermodynamical limit.

Obviously, the pieces such as OAB
(the letters refer to Figure~\ref{fig:fe2}), where all four curves are almost
coincident, can be expected to
form part of the true critical manifold. In particular we note that all curves
go {\em exactly} through the point
B with coordinates $(q,v)=(0,-4)$. By duality, and using results of
\cite{JSS05}, we can therefore deduce that 
spanning forests on the dual (union-jack) lattice undergo a phase transition
when the weight of each
component tree is $w_c = -4$.

Other parts of the curves, such as the stretched-out bubbles extending from B to
C, only emerge for sufficiently
large $n$ (here, for $n \ge 3$). This is true as well for the vertical rays that
build up at G, H and I. In analogy with the
square-lattice case, we expect the first two rays (at G and H) to be have $q$
coordinate $B_4 = 2$
and $B_6 = 3$. There is good evidence for conjecturing that the last ray (at I)
is situated at
$B_8 = 2 + \sqrt{2} = 3.414\,213\cdots$. The prong advancing at F seems to close
up the space
between the G and H vertical rays. Similarly, the prong advancing at E makes it
plausible that in the
$n \to \infty$ limit the ``upper curve'' OFE will extend to $q=4$, turn around,
and join the ``lower curve''
containing BC.

{}From these pieces of information we arrive at the following expectations---or
conjectures---for the critical manifold in
the continuum limit. The curve OAB and the ferromagnetic critical curve will
remain. Other curves will
extend to infinity in the antiferromagnetic regime, such as the one marked D in
Figure~\ref{fig:fe2}, and
possibly another emanating from E. The Berker-Kadanoff phase will be bordered by
the curve OFEPCB, where
P is a point with $q=4$. Inside this phase there will be vertical rays at
$q=B_4, B_6, B_8, B_{10},\ldots$ (the
figure provides good evidence for the first three rays at letters G, H and I).
Moreover, by analogy, we conjecture
that the square-lattice model will have the same infinite set of rays (although
Figure~\ref{fig:sq1} only provides
evidence for the first two of them).

\subsection{Kagome lattice}

In the case of the kagome lattice we have more information, since $P_B(q,v)$ has
been computed with
both square and hexagonal bases.%
\footnote{The results for square bases with $n \le 2$ have previously appeared
in \cite{Jacobsen12}.}
On the other hand, the phase diagram is more complicated. As already
pointed out in \cite{Jacobsen12}, no basis---however big---is likely to reveal
all aspects of the critical manifold,
and the general picture can only be understood by carefully comparing the
results from different bases and
embeddings.

\begin{figure}
\begin{center}
\includegraphics[width=12cm]{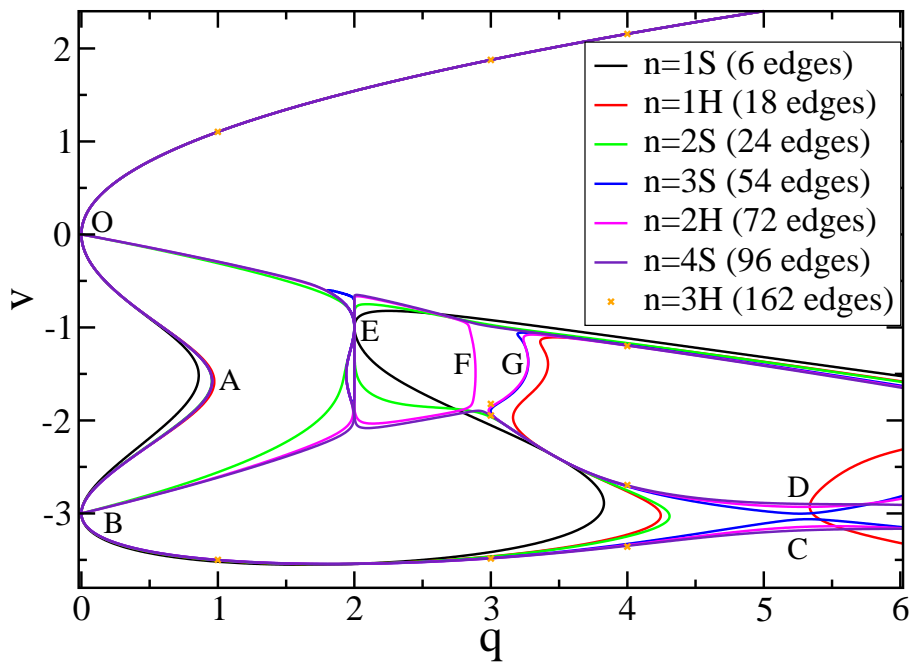}
\caption{Roots of $P_B(q,v)$ for the Potts model on the kagome lattice. In the
figure legend, the square bases are
 labelled $n$S, while $n$H denote the hexagonal bases.}
\label{fig:kag1}
\end{center}
\end{figure}

The roots of $P_B(q,v)$ are shown in Figure~\ref{fig:kag1}. The continuation of
the ferromagnetic critical curve
goes through OAB, through the point C, and out to infinity. Two other branches
go to infinity in the antiferromagnetic
region: one labelled D and another emanating from G. Notice that the two
branches C and D are not visible
with the smaller bases, since the curves join up and turn around at $q \approx
4$.

\begin{figure}
\begin{center}
\includegraphics[width=12cm]{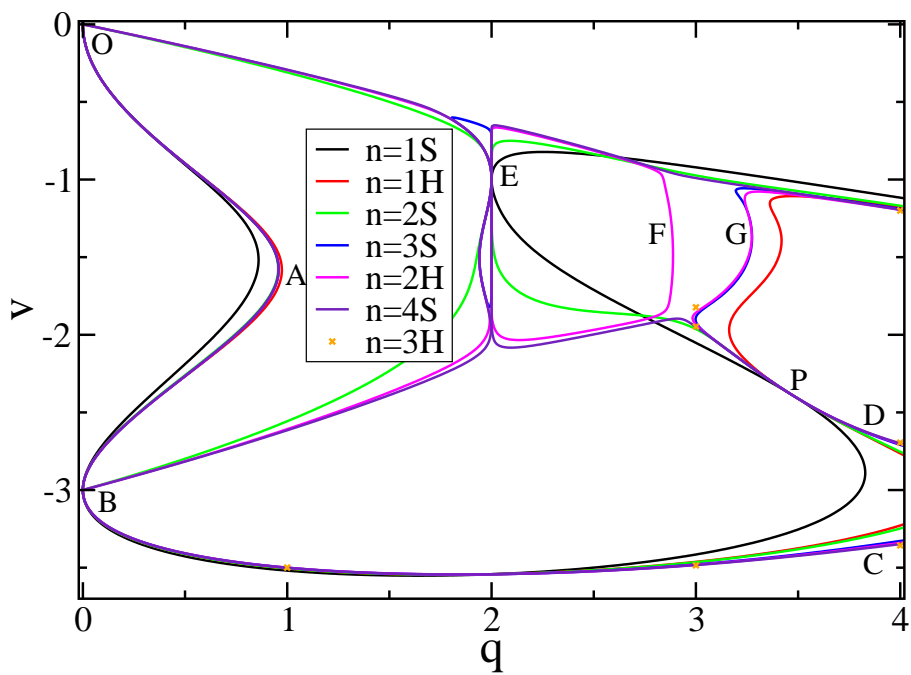}
\caption{Close-up on the antiferromagnetic region of Figure~\ref{fig:kag1}.}
\label{fig:kag2}
\end{center}
\end{figure}

The antiferromagnetic region contains interesting new features, as shown in the
close-up in Figure~\ref{fig:kag2}.
As before, there are vertical rays developing at $q=B_4$ (labelled E) and at
$q=B_6$ (labelled F).
Notice that the latter ray (F) is only revealed by the $n=2$ hexagonal basis;
this nicely illustrates
the remark made in the first paragraph of this subsection. There also seems to
be parity effects in the
size $n$ of the square bases. For instance, although the segment near G is
present in both hexagonal bases,
and in the $n=3$ square basis, it is absent from the larger $n=4$ square basis.
The fact that the $n=3$ square
and $n=2$ hexagonal curves are almost coincident near G makes it likely that
this segment will not move
much further upon increasing $n$. In particular, we note that this segment is
unlikely to become a vertical ray,
and therefore presumably is the rightmost termination of the Berker-Kadanoff
phase. If so, there will be only
two vertical rays (at $B_4$ and $B_6$) for this lattice.

The vertical extent of the rays at E and F reveals that the Berker-Kadanoff
phase is not bounded from below
by BC, but rather by another curve that emerges from B with positive slope.

\begin{figure}
\begin{center}
\includegraphics[width=12cm]{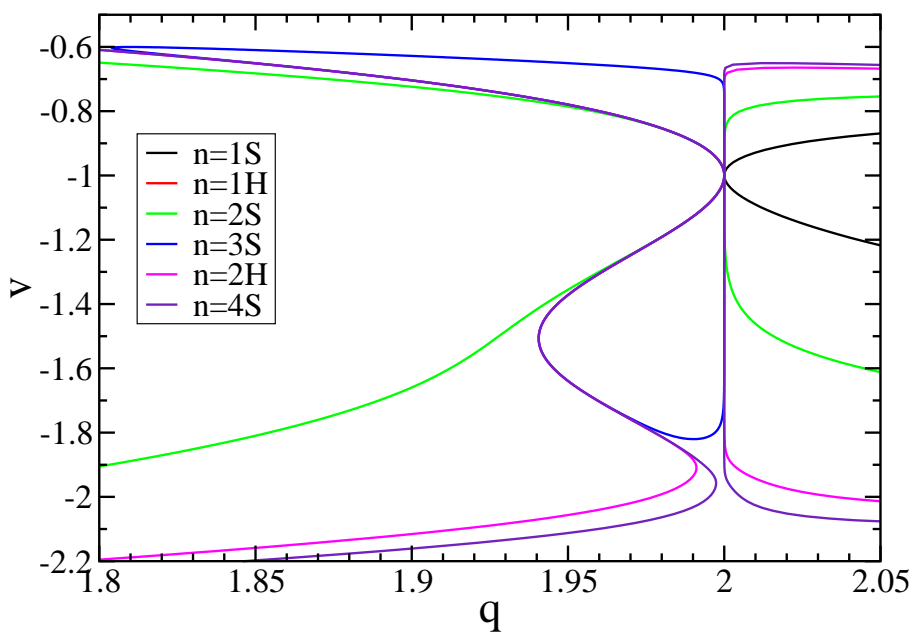}
\caption{Tiny detail of Figure~\ref{fig:kag1} in the region $q \approx 2$.}
\label{fig:kag3}
\end{center}
\end{figure}

Further magnification, shown in Figure~\ref{fig:kag3}, unearths an interesting
detail in the region $q \approx 2$.
Indeed, there is an ``unexpected curve'' emanating from $(q,v) = (2,-1)$,
exactly for any $n$, that goes to the left
through a point $\approx (1.94,-1.5)$ and ends near $(2,-2)$. On the scale of
Figure~\ref{fig:kag1} this produces a
very narrow sliver that one would be likely to dismiss as a finite-size effect.
But on the scale of Figure~\ref{fig:kag3}
it becomes clear that the different bases produce almost coincident results for
the location of the unexpected curve.
We therefore believe that this unexpected curve is a real effect that will
persist in the thermodynamical limit.

We remark that in the thermodynamical limit the critical manifold should contain
the point $(q,v) = (3,-1)$ exactly. Indeed, one
can show \cite{MooreNewman00} that the three-state zero-temperature
antiferromagnet on the kagome lattice is equivalent to the
corresponding four-state model on the triangular lattice. The latter
is known to be critical with central charge $c=2$ (see
\cite{MooreNewman00} and references therein). Our finite bases locate
the antiferromagnetic transition in the $q=3$ model at
\be
 v_{\rm c}^{\rm AF}(q=3) = \left \lbrace
 \begin{array}{lll}
 -0.921\,400\,117\cdots & \mbox{(6-edge basis)} \\
 -0.973\,665\,377\cdots & \mbox{(24-edge basis)} \\
 -0.990\,228\,473\cdots & \mbox{(96-edge basis)} \\
  \end{array} \right.
  \label{eq:q3v-1kagome}
\ee
and it seems likely that this might indeed tend to $v_{\rm c}^{\rm AF}(q=3) =
-1$
in the thermodynamical limit.

Note finally that all curves pass through $(q,v)=(0,-3)$ exactly.
Using again \cite{JSS05}, this implies that on the dual (diced) lattice, the
problem of
spanning forests \cite{JSS05} has a critical point with a
weight per tree $w_{\rm c} = -3$.

\subsubsection{A peculiar critical point}

Close inspection of Figure~\ref{fig:kag2} reveals that the curves for any basis---be it square or
hexagonal---go through the common point P with coordinates $(q,v) \approx (3.477,-2.393)$. This is strong evidence
that P might actually be an exact critical point for the kagome-lattice Potts model. We now
show that this is indeed the case, and we determine the coordinates and universality class of  P exactly.

\begin{figure}
\begin{center}
\includegraphics{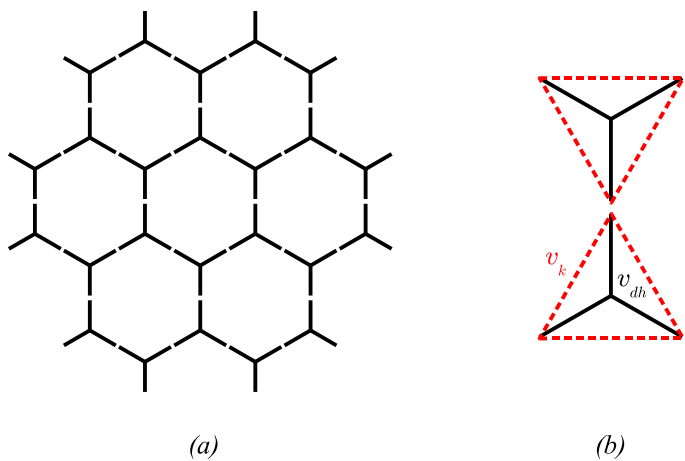}
\caption{a) The hexagonal lattice with doubled edges; b) a star-triangle replacement that gives the kagome lattice.}
\label{fig:doublehex}
\end{center}
\end{figure}

The crux of the argument is to relate the kagome lattice to a decorated hexagonal lattice, by means
of a star-triangle transformation. This is illustrated in Figure~\ref{fig:doublehex}. The latter lattice can
in turn be transformed into a standard hexagonal lattice, for which the exact critical curve is given by
(\ref{hex_latt_cc}).

Let us denote by $K_{\rm k}$ and $K_{\rm dh}$ the couplings between neighbouring Potts spins 
on the kagome and decorated
hexagonal lattice, respectively. Assume initially $q$ integer; the argument eventually carries over to
non-integer $q$ by analytical continuation. The star-triangle transformation then reads
\be
 \sum_{\sigma_0=1}^q {\rm e}^{K_{\rm dh} (\delta_{\sigma_1,\sigma_0} +
  \delta_{\sigma_2,\sigma_0} + \delta_{\sigma_3,\sigma_0})}
 = A {\rm e}^{K_{\rm k} (\delta_{\sigma_1,\sigma_2} + \delta_{\sigma_2,\sigma_3} + \delta_{\sigma_3,\sigma_1})} \,,
 \label{eq:star-triangle}
\ee
where $\sigma_1,\sigma_2,\sigma_3$ denote the three exterior spins common to a triangle and its inscribed star;
$\sigma_0$ is the internal spin of the triangle; and $A$ is a proportionality constant to be determined.
The relation (\ref{eq:star-triangle}) must hold for any choice of the exterior spins. The three possible
cases --- 1) $\sigma_1=\sigma_2=\sigma_3$; 2) $\sigma_1 = \sigma_2 \neq \sigma_3$; and 3) all three spins
different --- lead to the equations
\begin{eqnarray}
 {\rm e}^{3 K_{\rm dh}} + (q-1) &=& A {\rm e}^{3 K_{\rm k}} \,, \label{eq:st1} \\
 {\rm e}^{2 K_{\rm dh}} + \rm e^{K_{\rm dh}} + (q-2) &=& A {\rm e}^{K_{\rm k}} \,, \label{eq:st2} \\
 3 {\rm e}^{K_{\rm dh}} + (q-3) &=& A \,. \label{eq:st3}
\end{eqnarray}
Using (\ref{eq:st3}) to eliminate $A$ from Eqs.~(\ref{eq:st1})--(\ref{eq:st2}), and trading the couplings $K$ for the
Fortuin-Kasteleyn variables $v = {\rm e}^K - 1$ as usual, we arrive at
\begin{eqnarray}
 (v_{\rm dh} + 1)^3 + q - 1 &=& \big( 3(v_{\rm dh}+1) + q - 3 \big) (v_{\rm k} + 1)^3 \,, \label{eq:dh1} \\
 (v_{\rm dh} + 1)^2 + (v_{\rm dh}+1) + q - 2 &=& \big( 3(v_{\rm dh}+1) + q - 3 \big) (v_{\rm k} + 1) \label{eq:dh2} \,.
\end{eqnarray}

The decorated hexagonal lattice can be transformed into a standard hexagonal lattice by applying the series reduction
formula \cite{Sokal05} to turn the double edges into simple edges. This reads
\be
 v_{\rm h} = \frac{v_{\rm dh}^2}{q + 2 v_{\rm dh}} \,. \label{eq:dh3}
\ee
The resulting hexagonal-lattice Potts model is critical when (\ref{hex_latt_cc}) is satisfied, that is
\be
 v_{\rm h}^3 - 3 q v_{\rm h} - q^2 = 0 \,. \label{eq:dh4}
\ee

There are two real solutions of the equations (\ref{eq:dh1})--(\ref{eq:dh4}). The first one reads
\be
 q = 2 \,, \qquad v_{\rm k} = -1 \,, \qquad v_{\rm h} = v_{\rm dh} = -2 \,.
\ee
Indeed all the kagome-lattice critical polynomials have a root at $(q,v_{\rm k}) = (2,-1)$.
But more interestingly, we have the real solution
\be
 q = 3.476\,950\,573\,042\,399 \cdots \,, \qquad v_{\rm k} = -2.392\,646\,781\,702\,640 \cdots \,,
\ee
explaining the point P. Note that this corresponds to $v_{\rm dh} \approx -1.453$ and $v_{\rm h} \approx 3.701$,
so the equivalent coupling on the hexagonal lattice is positive. It is well-known from conformal field theory and
exact solutions that the ferromagnetic phase transition on any lattice with $q = \big( 2 \cos(\pi e_0) \big)^2$ is
second order with central charge $c = 1 -  \frac{e_0^2}{1 - e_0}$. The universality class of the transition
at point P is therefore characterised by
\be
 c = 0.905\,667\,155\,343\,907\cdots \,.
\ee

\subsection{$(3,12^2)$ lattice}

\begin{figure}
\begin{center}
\includegraphics[width=12cm]{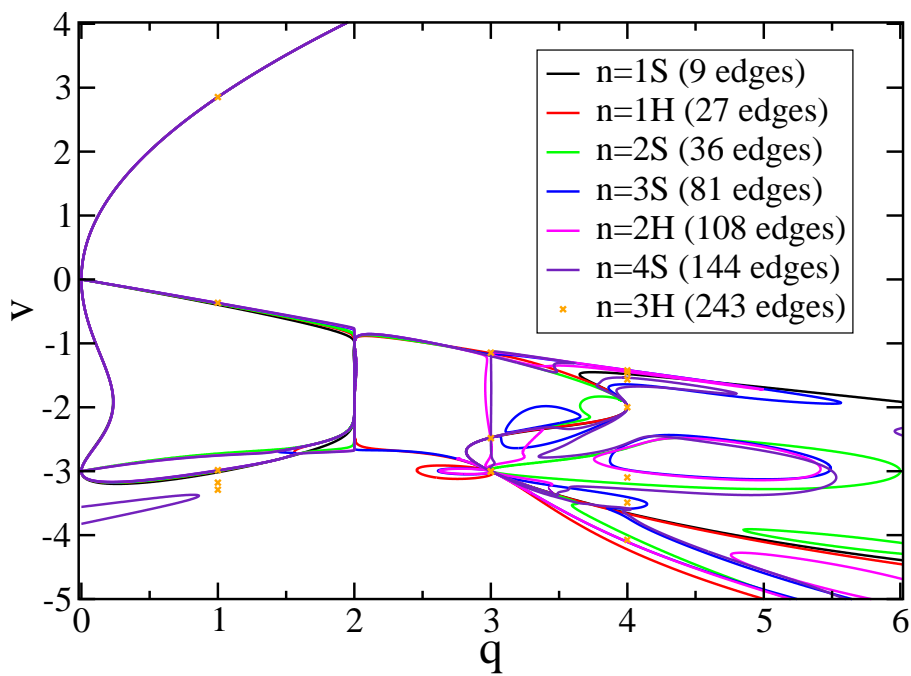}
\caption{Roots of $P_B(q,v)$ for the Potts model on the $(3,12^2)$ lattice. The
square (resp.\ hexagonal) bases
 of size $n$ are labelled $n$S (resp.\ $n$H).}
\label{fig:tt1}
\end{center}
\end{figure}

\begin{figure}
\begin{center}
\includegraphics[width=12cm]{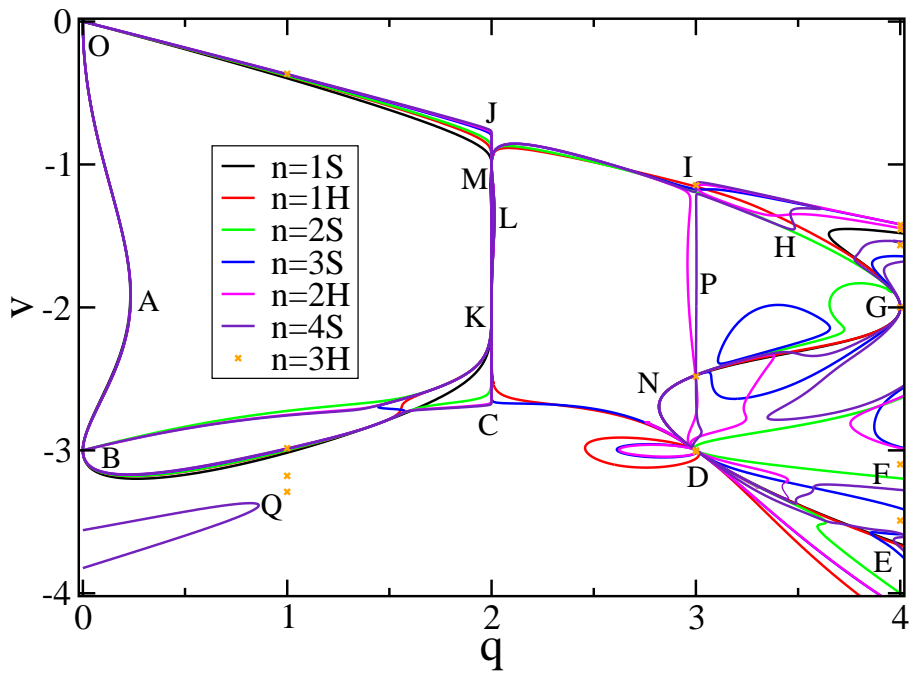}
\caption{Close-up on the antiferromagnetic region of Figure~\ref{fig:tt1}. The
various letters are referred to
in the main text.}
\label{fig:tt2}
\end{center}
\end{figure}

The critical manifold obtained from studying the roots of $P_B(q,v)$ on the
$(3,12^2)$ lattice
is shown in Figure~\ref{fig:tt1}. This is considerably more involved than for
the other lattices
studied this far. A close-up on the antiferromagnetic region is depicted in
Figure~\ref{fig:tt2}.

The curve OAB is well-converged as usual. As in the kagome case, there are two
curves originating
from B, namely BK and BC, but unlike the kagome case both now end on the
vertical ray CJ at $q=B_4$.
Another prong Q, visible only from the largest basis, might eventually provide
further curves going towards
C, or beyond.

Note that the point B is at $(q,v)=(0,-3)$ exactly, implying \cite{JSS05} that
spanning forests on the dual
(asanoha, or hemp leaf) lattice have a critical fugacity $w_c = -3$ per
component tree.

The lower boundary of the Berker-Kadanoff phase, BC, goes on via CD, where it
encounters another vertical
ray P at $q=B_6$, extending between D and I.
There are some horizontally elongated ``bubbles'' near D, but since their size
decreases
with $n$ it is uncertain whether they persist in the thermodynamical limit.
Meanwhile, the upper boundary
of the Berker-Kadanoff phase starts out as OJI. The region $3 \le q \le 4$ is
particularly complicated and will
be discussed further below.

\begin{figure}
\begin{center}
\includegraphics[width=12cm]{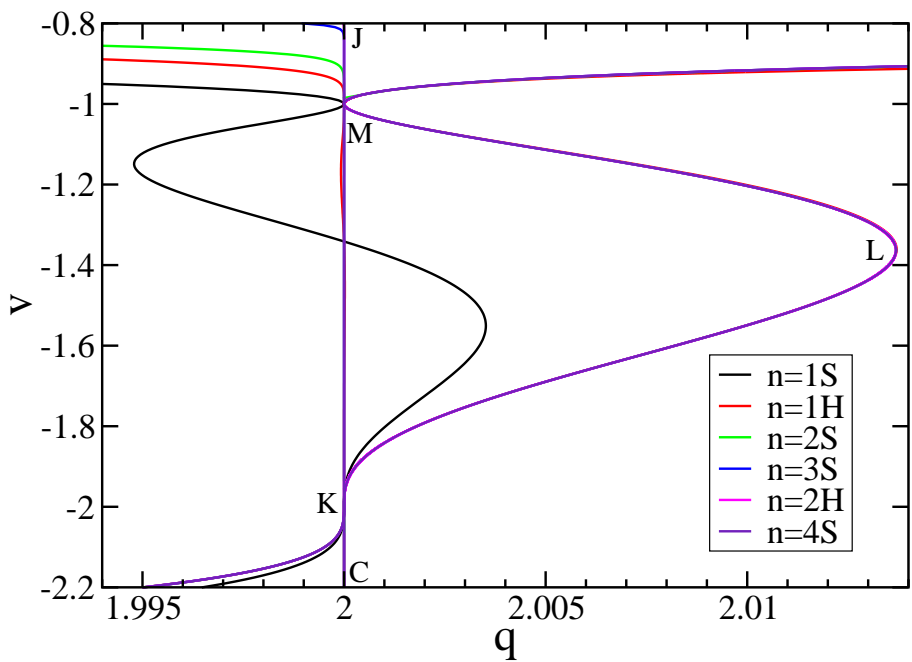}
\caption{Tiny detail of Figure~\ref{fig:tt1} in the region $q \approx 2$.}
\label{fig:tt3}
\end{center}
\end{figure}

Like for the kagome lattice, there are interesting details in the region $q
\approx 2$.
This is shown in magnification in Figure~\ref{fig:tt3}.
There is again an ``unexpected curve'' emanating from point M with coordinates
$(q,v) = (2,-1)$, exactly for any $n$,
that goes now to the right through point L with $(q,v) \approx (2.014,-1.35)$, for all but the smallest ($n=1$ square) basis,
and ends at point K with
$(q,v) = (2,-2)$, again exactly for any $n$.
Once again, the agreement between the largest bases is such that we can believe
that this unexpected curve
will subsist in the thermodynamical limit.

\begin{figure}
\begin{center}
\includegraphics[width=12cm]{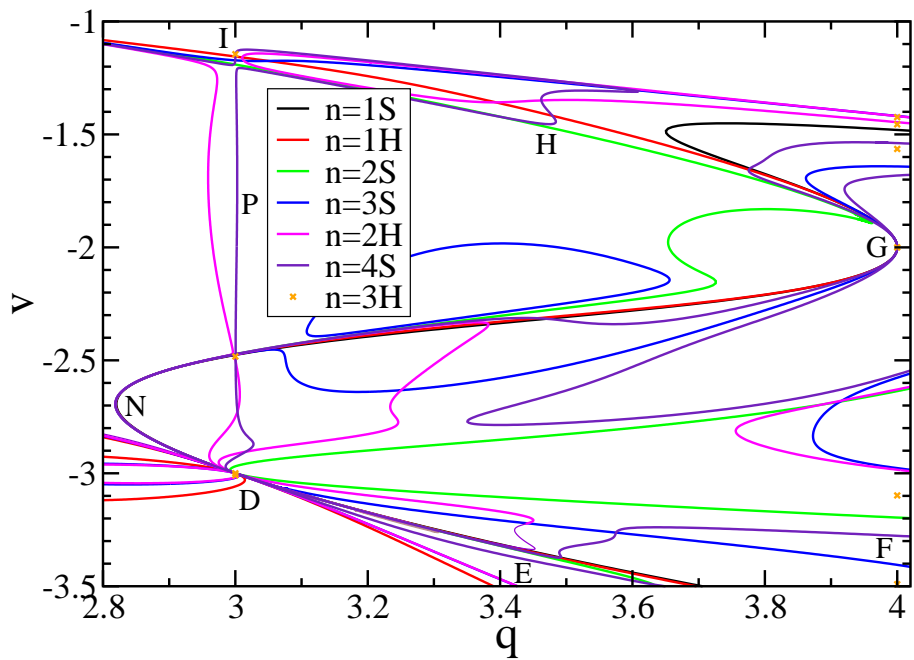}
\caption{Magnification of the region with $2.8 \le q \le 4$ in
Figure~\ref{fig:tt1}.}
\label{fig:tt4}
\end{center}
\end{figure}

The most complicated region is shown enlarged in Figure~\ref{fig:tt4}. The
boundary of the Berker-Kadanoff phase
might be given by DGHI. Note that the point G is $(q,v)=(4,-2)$, exactly for all
the biggest sizes.
Several of the curves contain ``wrinkles'' or other signatures close to $q = B_8
= 3.414\,213\cdots$, such as 
the one labelled H. We take this as a sign of an emergent vertical ray at $B_8$.
In conjunction with the fact that
point G has $q = B_\infty = 4$, this leads us to believe that the
thermodynamical limit will in fact have vertical
rays at all $B_{2k}$ with $k \ge 2$.

We further remark that many curves pass through $(q,v) = (3,-3)$ exactly. By
duality, this means that the
$3$-state Potts antiferromagnet on the dual (asanoha) lattice should undergo a
phase transition
at zero temperature.

The curve DNG inside the Berker-Kadanoff phase should also be noticed. Finally,
there are several
curves going from D towards infinity, such as those marked E and F in
Figure~\ref{fig:tt2}.

\section{Discussion}
\label{sec:disc}

In this work, we have given a new definition of the critical polynomial for the $q$-state Potts model that we had previously defined by the contraction-deletion identity \cite{Jacobsen12}. This has allowed us to compute these polynomials for various lattices using the transfer matrix, a method that permits the use of much larger bases, and therefore the calculation of much higher-degree polynomials, than the contraction-deletion algorithm. Our results put beyond doubt the conjecture that critical polynomials, when they do not provide the exact critical frontier, give excellent approximations that approach the exact answer in the limit of infinite bases. In the ferromagnetic regime, we were able to locate critical couplings on the kagome, $(4,8^2)$ and $(3,12^2)$ lattices for $q=3$ and $q=4$ with accuracy rivaling or exceeding that of traditional Monte Carlo or transfer matrix diagonalisation methods. Moreover, the polynomial estimates for $q=3$ are comparable in precision to those for $q=4$, and thus appear not to suffer from the logarithmic corrections to scaling that plague standard numerical techniques for $q=4$.

Critical polynomials also give a clear look at the antiferromagnetic region of the phase diagram, including the Berker-Kadanoff phase, which had previously been difficult to observe numerically. For the square lattice, we find predictions that are completely consistent with theoretical understanding of the BK regime, with vertical rays located at the Beraha numbers $B_4$ and $B_6$. For the other lattices studied here, on which less is known about the antiferromagnetic region, we find qualitatively similar behaviour, but with notable differences. On the kagome lattice, we observed a previously unknown point in the AF region that the polynomials for every basis placed on the critical curve. Given this information, we were able to find an argument that established this as an exact critical point of the kagome Potts model by a transformation from a decorated hexagonal lattice. This demonstrates the power of the critical polynomial method beyond the numerical determination of critical points --- parameters of potential exact solutions are prominently displayed in the phase diagram. It seems likely that there remain many others to be found in this way. Similarly, on the kagome and $(3,12^2)$ lattices, we have found unexpected critical curves within the Berker-Kadanoff phase that are predicted by a range of bases, indicating that they very likely represent real features of the phase diagrams. The presence of these curves awaits a theoretical explanation.

Several factors determine the accuracy of the predictions made by the critical polynomial method. Generally speaking, the best accuracy is
obtained in the ferromagnetic region ($v > 0$). On the other hand, we have observed for all lattices studied here that the critical points of the
Ising model ($q=2$) come out exactly, even when they are situated in the antiferromagnetic region ($v < 0$). A similar phenomenon holds true
for $q=0$, insofar as all the curves pass exactly through the origin $(q,v)=(0,0)$ as well as through another point $(0,v)$, with $v=-4$ for the square
and $(4,8^2)$ lattices, and $v=-3$ for the kagome and $(3,12^2)$ lattices. We have seen that other exact points may appear as well on certain
lattices. The limits $q \to \infty$ can also easily be shown to be exact, in the sense that the critical polynomial provides the correct asymptotic
behaviour as predicted by first-order phase coexistence in the Fortuin-Kasteleyn expansion \cite{Jacobsen12}.
 Outside these exact points and limits, the accuracy obviously depends on the size of the basis, and on the compatibility of its embedding with the
symmetries of the lattice. For example, hexagonal bases fare better than their square counterparts when the lattice has a 3-fold rotational
symmetry. For square bases the twist also seems to play a role, with the untwisted bases being the most accurate for the kagome and $(3,12^2)$ lattices,
and by contrast, the maximally twisted bases performing better for the $(4,8^2)$ lattice. Finally, as an empirical rule, it appears that the accuracy
deteriorates with increasing distance along the curve from one of the exact cases. A good example of that point is the comparatively mediocre precision
with which the method accounts for the known $(q,v)=(3,-1)$ critical point of the kagome-lattice Potts model; see (\ref{eq:q3v-1kagome}).

Because this method is relatively new, the ultimate limit on the size of basis that can be employed on each lattice is not yet completely clear. Here, we were able to find polynomials of degree up to 243 using a large parallel calculation. However, improvements are certainly possible and it will hopefully come to be seen as a worthy computational challenge to push this limit even further. Aside from optimising performance, there remains a great deal to be explained about the critical polynomial method. The fact that it works so well at predicting unsolved critical manifolds is still quite mysterious. Although one can argue that universality guarantees equation (\ref{eq:2D0D}) will give estimates that approach the correct value for infinite $B$, it is surprising how accurate the results are for small bases. It is clear that the condition (\ref{eq:2D0D}) reveals some larger truth about critical Potts systems, the full mathematical implications of which are yet to be discovered.

\section*{Acknowledgments}

The work of JLJ was supported by the Agence Nationale de la Recherche
(grant ANR-10-BLAN-0414:~DIME) and the Institut Universitaire de
France. We thank John Cardy for a valuable suggestion concerning the equivalence of the polynomial definitions 
and Jim Glosli at LLNL for helpful advice on the parallel implementation of the transfer matrix code. Additionally, CRS
wishes to thank Bob Ziff for discussions and collaboration on related
work. We are grateful to the
Mathematical Sciences Research
Institute at the University of California, Berkeley for hospitality
during the programme on Random Spatial Processes where this work was
initiated. CRS also thanks the Institute for Pure and Applied Mathematics at
UCLA, where part of this work was performed.

\section*{References}
\bibliographystyle{iopart-num}
\bibliography{SJ}

\end{document}